\documentclass[fleqn,usenatbib]{mnras}

\usepackage{newtxtext,newtxmath}

\usepackage[T1]{fontenc}

\DeclareRobustCommand{\VAN}[3]{#2}
\let\VANthebibliography\thebibliography
\def\thebibliography{\DeclareRobustCommand{\VAN}[3]{##3}\VANthebibliography}

\usepackage{graphicx}	
\usepackage{amsmath}	
\usepackage{comment}

\newcommand{\kms}{\mbox{km~s$^{-1}$}}
\newcommand{\Msun}{\mbox{$\text{M}_{\sun}$}}
\newcommand{\dco}{\mbox{C$^{18}$O}}
\newcommand{\mm}{Mystic Mountains}

\title[the MUSE view of the Mystic Mountains]{Into the Mystic: the MUSE view of the ionized gas in the Mystic Mountains in Carina}

\author[M. Reiter et al.]{
Megan Reiter,$^{1}$\thanks{E-mail: megan.reiter@rice.edu (MR)}
Anna F. McLeod,$^{2,3}$
Dominika Itrich$^{4,5}$
and Pamela D. Klaassen$^{6}$
\\
$^{1}$Department of Physics and Astronomy, Rice University, 6100 Main St - MS 108, Houston, TX 77005, USA \\
$^{2}$Centre for Extragalactic Astronomy, Department of Physics, Durham University, South Road, Durham DH1 3LE, UK \\
$^{3}$Institute for Computational Cosmology, Department of Physics, University of Durham, South Road, Durham DH1 3LE, UK \\
$^{4}$European Southern Observatory, Karl-Schwarzchild-Str. 2, D-85748 Garching bei M\"{u}nchen, Germany \\ 
$^{5}$Steward Observatory, The University of Arizona, 933 N. Cherry Ave, Tucson, AZ 85721, USA \\  
$^{6}$UK Astronomy Technology Centre, ROE, Blackford Hill, Edinburgh, EH9 3HJ, UK 
}

\date{Accepted 2025 January 28. Received 2024 December 23; in original form 2024 November 1}

\pubyear{2025}

\begin{document}
\label{firstpage}
\pagerange{\pageref{firstpage}--\pageref{lastpage}}
\maketitle

\begin{abstract}
We present optical integral field unit (IFU) observations of the \mm, a dust pillar complex in the center of the Carina Nebula that is heavily irradiated by the nearby young massive cluster Trumpler~14. With the continuous spatial and spectral coverage of data from the Multi-Unit Spectroscopic Explorer (MUSE), we measure the physical properties in the ionized gas including the electron density and temperature, excitation, and ionization. MUSE also provides an excellent view of the famous jets HH~901, 902, and 1066, revealing them to be high-density, low-ionization outflows despite the harsh environment. 
HH~901 shows spatially extended [C~{\sc i}] emission tracing the rapid dissociation of the photoevaporating molecular outflow in this highly irradiated source. 
We compute the photoevaporation rate of the \mm\ and combine it with recent ALMA observations of the cold molecular gas to estimate the remaining lifetime of the \mm\ and the corresponding shielding time for the embedded protostars. The longest remaining lifetimes are for the smallest structures, suggesting that they have been compressed by ionizing feedback. Our data do not suggest that star formation in the \mm\ has been triggered but it does point to the role that ionization-driven compression may play in enhancing the shielding of embedded stars and disks. Planet formation models suggest that the shielding time is a strong determinant of the mass and orbital architecture of planets, making it important to quantify in high-mass regions like Carina that represent the type of environment where most stars form. 
\end{abstract}




\section{Introduction}

Most stars form in high-mass star-forming regions \citep[e.g.,][]{krumholz2019}.  
Energy and momentum from winds, radiation, and eventually supernova explosions of high-mass stars, collectively called feedback, are deposited back into the local environs and play a critical role in regulating the lifetime and efficiency of star-forming molecular clouds \citep[e.g.,][]{krumholz2019,kruijssen2019,chevance2022,chevance2023}. 
Radiation from the high-mass stars provides most of the star-formation rate tracers used in other galaxies -- either the direct (UV) radiation from the high-mass stars themselves \citep[e.g.,][]{calzetti2015,adamo2017,lee2022} or emission from atoms and molecules excited by it [e.g., H-recombination lines, and, in younger regions, emission from polycyclic aromatic hydrocarbons (PAHs), e.g.,  \citealt{belfiore2023,sandstrom2023_nircam,sandstrom2023_miri,scheuermann2023,calzetti2024,gregg2024}].
The global view afforded by observations of external galaxies reveals how feedback varies with age and environment \citep[e.g.,][]{lopez2011,lopez2014,mcleod2019,mcleod2020,mcleod2021,barnes2021}.

Many studies of Milky Way clouds also focus on a global assessment of feedback as a function of time and environment \citep[e.g.,][]{schneider2020,barnes2020,olivier2021}. 
However, within the Galaxy, it is possible to resolve the impact of feedback as it profoundly reshapes the star-forming environment by 
heating gas and dust in the region \citep[e.g.,][]{roccatagliata2013,rebolledo2016}, 
carves channels for winds and hot gas to escape \citep[e.g.,][]{townsley2003,rosen2014,bonne2022}, 
compresses and accelerates gas away from the ionizing sources \citep[e.g.,][]{sandford1982,bertoldi1989,bertoldi1990,bertoldi1992,bisbas2011,esquivel2007,goicoechea2016,reiter2020_combined}, 
and may affect the chemistry of the star-forming gas through UV irradiation \citep[e.g.,][]{cuadrado2017,goicoechea2019,goicoechea2021,berne2022} and cosmic ray bombardment \citep[e.g.,][]{bisbas2017,aharonian2019,gabici2022,desch2022,krumholz2023,peron2024}. 
Resolving and connecting the feedback from high-mass stars to the physical conditions in nearby cold, molecular gas is increasingly of interest to the star- and planet-formation community as high-mass regions are now readily accessible with the advent of the \emph{James Webb Space Telescope} and the Atacama Large Millimeter Array (ALMA).

Dust pillars are one of the most famous and most prominent feedback-carved structures \citep[e.g., the famous `Pillars of Creation';][]{hester1996,linsky2007,ercolano2012,westmoquette2013,mcleod2015,pattle2018,xu2019_M16,sofue2020,karim2023,dewangan2024}.
It has been proposed that pillars reflect stimulated star formation where external irradiation triggers the formation of the stars frequently found at the pillar tips \citep[e.g.,][]{smi10b,ohl12,tremblin2013,paron2017,djupvik2017}. 
However, pre-existing over-densities can also shield cloud material behind them, leading to the formation of dust pillars. 
Often, it is difficult to distinguish between these two mechanisms \citep[e.g.,][]{bowler2009,alexander2013,dale2015}.

Simulations show that the shape of the radiation field and the underlying structure of the gas can affect whether feedback enhances or suppresses star formation
\citep[e.g.,][]{gritschneder2009,gritschneder2010,tremblin2012_sc,tremblin2012_sct,dale2012,walch2013,boneberg2015,dale2017,ali2018,menon2020,grudic2021}.
Instead of triggering collapse, momentum injected by photoevaporation may drive turbulence in the cold molecular gas, helping to support the cloud against collapse and fragmentation. 
Resolved observations are required to determine the balance of these two effects. 

The Carina Nebula is one of the nearest \citep[2.35~kpc,][]{goppl2022} regions of high-mass star-formation in the Galaxy.
With hundreds of O- and B-type stars, including some of the earliest spectral type stars in the Galaxy, Carina samples conditions that more closely resemble extra-galactic star-forming regions. 
Carina is a `cluster of clusters', with the famous young massive clusters Trumpler~14 and 16 (Tr~14 and Tr~16, respectively) at its heart. 
The older Tr15 cluster lies in the northern part of the complex while the most active star formation in the region is in the South Pillars where thousands of young stellar objects (YSOs) are found in feedback-carved dust pillars \citep[e.g.,][]{megeath1996,rathborne2004,smi10b,tapia2011,gaczkowski2013,tung2024}.

Targeted observations of pillars in Carina and other high-mass star-forming regions provided the first quantitative measure of how ionized gas conditions change as a function of local UV irradiation \citep{mcleod2016}. 
With ALMA, it is now possible to resolve the structure and kinematics of the cold molecular gas in these same regions \citep{klaassen2020}. 
It is only with spatially and spectrally resolved observations that we can begin to distinguish between pillar formation mechanisms. 
For example, an analysis of the molecular gas in six pillars in Carina by \citet{menon2021} found predominantly compressive modes of turbulence in the pillars that may enhance the star formation rate.

Fewer studies target the highly irradiated central region of Carina. 
Two groups observed a heavily irradiated dust wall illuminated by Tr14 \citep{rebolledo2020,hartigan2022} but come to conflicting conclusions about the impact that feedback has had on the cold molecular gas. 
\citet{rebolledo2020} find fewer but more massive cores, consistent with turbulent fragmentation, as compared to a less irradiated cloud in the South Pillars. 
\citet{downes2023} argue that there is no evidence for large-scale turbulence driving from the external irradiation. 
In this picture, pillars represent a different set of conditions than dust walls because they are the result of compressive turbulence.
Resolving the tensions between these results requires sampling a wider range of environments, including dust pillars in the most heavily irradiated parts of the cloud.

\begin{figure*}
	\includegraphics[width=\textwidth]{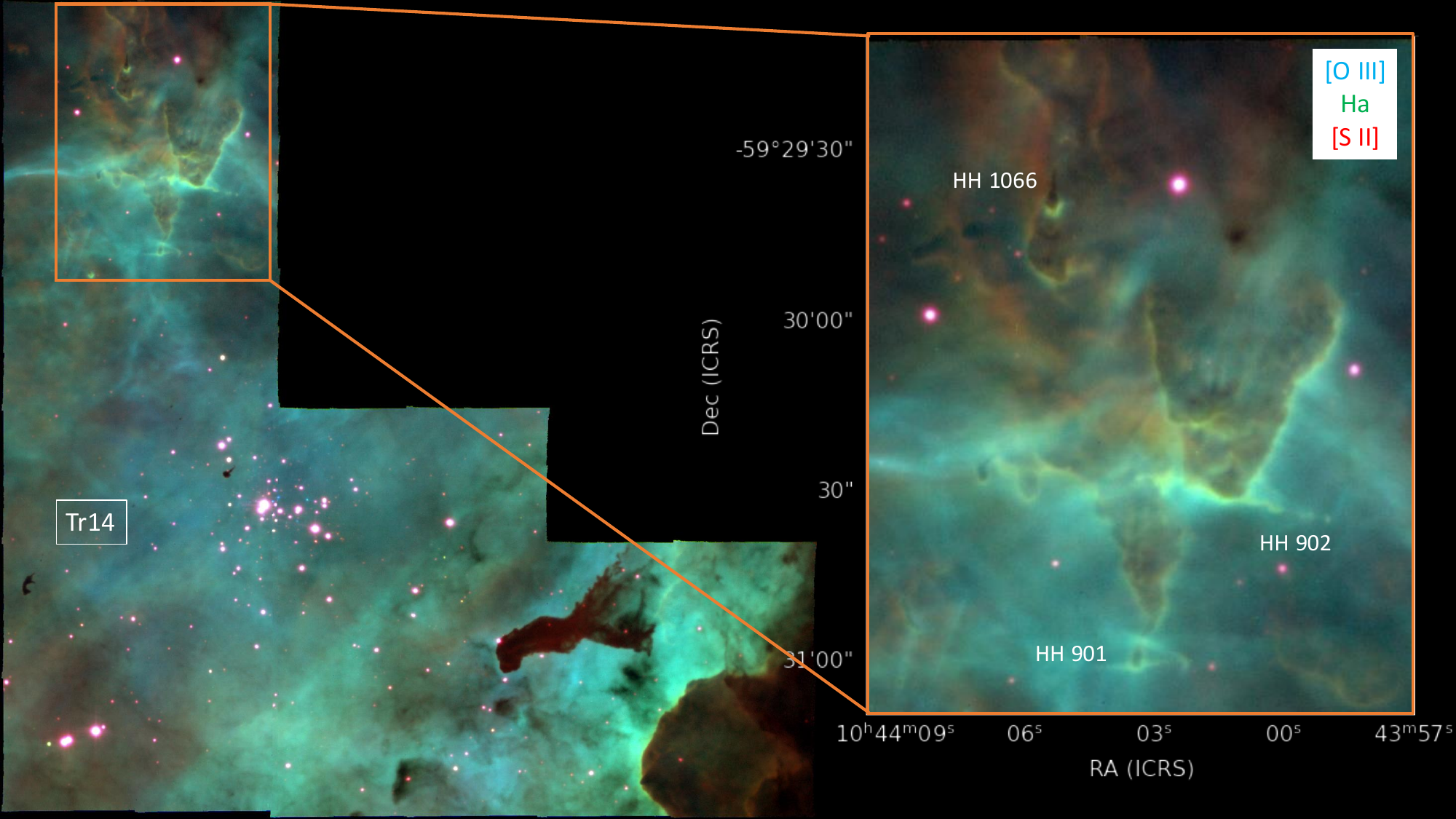}
    \caption{Color image showing the entire Tr14 MUSE survey area (north is up, east is to the left). The \mm\ are $\sim 1$~pc to the north (in projection) of the Tr14 young massive cluster. The zoomed in image shown to the right highlights the subset of the data used in this work with the famous jets HH~901, HH~902, and HH~1066 labelled. 
    }
    \label{fig:MM_intro}
\end{figure*}

In this paper, we target the so-called Mystic Mountains \citep[as they were called for the \emph{Hubble Space Telescope} 20$^{\mathrm{th}}$ anniversary image; Area 29 in][see Figure~\ref{fig:MM_intro}]{hartigan2015}. 
The heavily irradiated cloud complex lies $\sim 1$~pc north of Tr14 which illuminates the \mm\ with $Q_H \sim 10^{50}$~s$^{-1}$ ionizing photons. 
Three famous Herbig–Haro (HH) jets emerge from the tips of the pillars within the \mm\ -- HH~901, HH~902, and HH~1066 \citep[see Figure~\ref{fig:MM_intro} and][]{smith2010,reiter2013,reiter2014,cortes-rangel2020}. 
\citet{reiter2023_MM} recently looked at the large-scale gas kinematics in the \mm\ with modest spatial resolution  (6\arcsec\ beam). 
Gas properties are consistent with compressive turbulence with similar parameters to the less heavily irradiated pillars in \citet{menon2021}.

In this paper, we examine the ionized gas properties of the \mm\ using the broad spatial and spectral coverage of 
the Multi-Unit Spectroscopic Explorer \citep[MUSE;][]{bacon2010} on the European Southern Observatory's Very Large Telescope. 
Within the \mm, we sample a range of feedback conditions and trace the variations from `tip to tail' (corresponding to a factor of $\sim 2$ change in incident UV flux). 
By comparing with ALMA observations, our goal is to connect the dots between the input radiation and its consequences for the cold gas kinematics \citep[e.g.,][]{reiter2019_tadpole,reiter2020_alma,reiter2020_combined}.

\section{Observations}\label{s:obs}

The observations used in this paper are part of a larger survey of Tr14 and the surrounding nebulosity with MUSE (programme ID 097.C-0137; PI: A.~F.~McLeod; see also \citealt{itrich2024}). 
We use a subset of four MUSE cubes (see Figure~\ref{fig:MM_intro}) that capture the \mm. 
Data were obtained in Feb 2016 in service mode. 
Observations utilize wide-field mode observations which provide 1\arcmin\ $\times$ 1\arcmin\ field-of-view and continuous spectral coverage from $\lambda 4650-9300$~\AA. 
Spectral resolution varies from $R=2000-4000$ over this range. 

At each pointing, three separate exposures were obtained with a 90$^{\circ}$ rotation between each to minimize detector artifacts. 
To detect both bright and faint stars and nebulosity, each position was observed twice -- once with short (5~s) exposures and once with long (780~s) exposures. 
While the mosaic setup was the same for the short and long exposures, small offsets in the pointing led to slightly different field centers (see Table~\ref{t:obs}). 
The total integration time per pointing for the long exposures is 39~min and 15~s for the short exposures. 

Data were reduced using the MUSE recipes v.2.8.9 \citep{weilbacher2012} in the EsoRex environment with standard calibrations. 
We also use standard post-processing for all steps except for sky subtraction. 
In a standard reduction, the pipeline uses the darkest $\sim5$\% of pixels to estimate the telluric emission. 
However, in bright H~{\sc ii} regions like Carina, even these `dark' pixels have significant nebular emission. 
Important nebular lines have similar wavelengths to bright sky lines which can lead to them being oversubtracted in the automatic pipeline reduction. 
To correct for this, we use the modified sky subtraction approach developed by \citet{zeidler2019}. 
Briefly, this method uses a five-step process to remove telluric emission. 
First, the continuum and line emission are determined from the {\tt muse\_create\_sky} recipe. 
Second, the continuum emission is set to zero to avoid subtracting the continuum of the H~{\sc ii} region. 
Third, the {\tt muse\_create\_sky} recipe is run a second time to get an accurate flux estimate for the sky lines. 
Fourth, we assume that OH and O$_2$ lines dominate the telluric emission; all other telluric line fluxes are set to zero. 
Finally, this sky model is subtracted from the science observations during post-processing with the {\tt muse\_scipost} recipe.
To remove the continuum emission from the reduced data, we fit a fourth-order polynomial to a subset of pixels without strongly structured nebular emission and subtract this from the entire cube. 

To align each MUSE cube with \emph{Gaia}, we used the {\sc astrometry} python package from \citet{wenzl2022}. 
We extract an image from each cube to identify stars in the \emph{Gaia} catalog and compute the updated astrometric solution, then apply this solution to each MUSE cube.

For the majority of the analysis in this paper, we use only the long exposures as these provide the best signal-to-noise for most lines. 
However, H$\alpha$ emission is saturated in some portions of the long exposures, particularly along the ionization fronts. 
This is most pronounced in and around the HH~901 and HH~902 jets. 
For analysis of the H$\alpha$ line in these regions, we use only the short exposures (H$\alpha$ emission is not saturated in or near HH~1066, so we use the long exposures for that portion of the analysis). 
For line ratios computed with H$\alpha$, for example the H$\alpha$/H$\beta$ ratio in Section~\ref{ss:extinction}, we use the short exposures for both lines over the full map. 
Throughout the paper, we state explicitly when the short exposures are used.

\begin{table}
	\centering
	\caption{MUSE observations of the \mm\ in Carina. (1) is the pointing number; (2) is the R.A. in hh:mm:ss.s; (3) is the Dec. in $\pm$dd:mm:ss; (4) is the date of observation; and (5) is the exposure time in seconds.} 
	\label{t:obs}
	\begin{tabular}{lcccrr} 
		\hline\hline
		tile & RA (J2000) & Dec. (J2000) &  date & t$_{\mathrm{int}}$ [s] & seeing$^*$ \\
		\hline
        \multicolumn{6}{c}{long} \\
        \hline
		1 & 10:44:08.4 & $-$59:29:40.0 & 2016-02-28 & 2340 & 0.75\arcsec\ \\
		2 & 10:44:00.7 & $-$59:29:39.5 & 2016-02-25 & 2340 & 0.92\arcsec\ \\
		3 & 10:44:00.7 & $-$59:30:39.5 & 2016-02-25 & 2340 & 0.85\arcsec\ \\
        4 & 10:44:08.3 & $-$59:30:39.7 & 2016-02-25 & 2340 & 0.86\arcsec\ \\
		\hline
        \multicolumn{6}{c}{short} \\
        \hline
		1 & 10:44:08.3 & $-$59:29:39.9 & 2016-02-25 & 15 & 0.71\arcsec\ \\
		2 & 10:44:00.7 & $-$59:29:40.0 & 2016-02-25 & 15 & 0.77\arcsec\ \\
		3 & 10:44:00.7 & $-$59:30:38.5 & 2016-02-25 & 15 & 0.77\arcsec\ \\
        4 & 10:44:08.4 & $-$59:30:38.4 & 2016-02-25 & 15 & 1.1\arcsec\ \\
		\hline
  \multicolumn{6}{l}{$^*$ long exposure values from Table~A.1 in \citet{itrich2024}}
	\end{tabular}
\end{table}

\section{Results}\label{s:results}

Four MUSE pointings capture the famous \mm\ providing continuous spectral coverage over the full optical range. The pillar structure is well resolved, as are the famous HH jets: HH~901, HH~902, and HH~1066. 
Each jet emerges from the head of a distinct dust pillar within the \mm\ complex. 
In the following sections, we derive the physical properties of the hot, ionized gas associated with the \mm.

\begin{figure*}
    \includegraphics[width=\columnwidth]{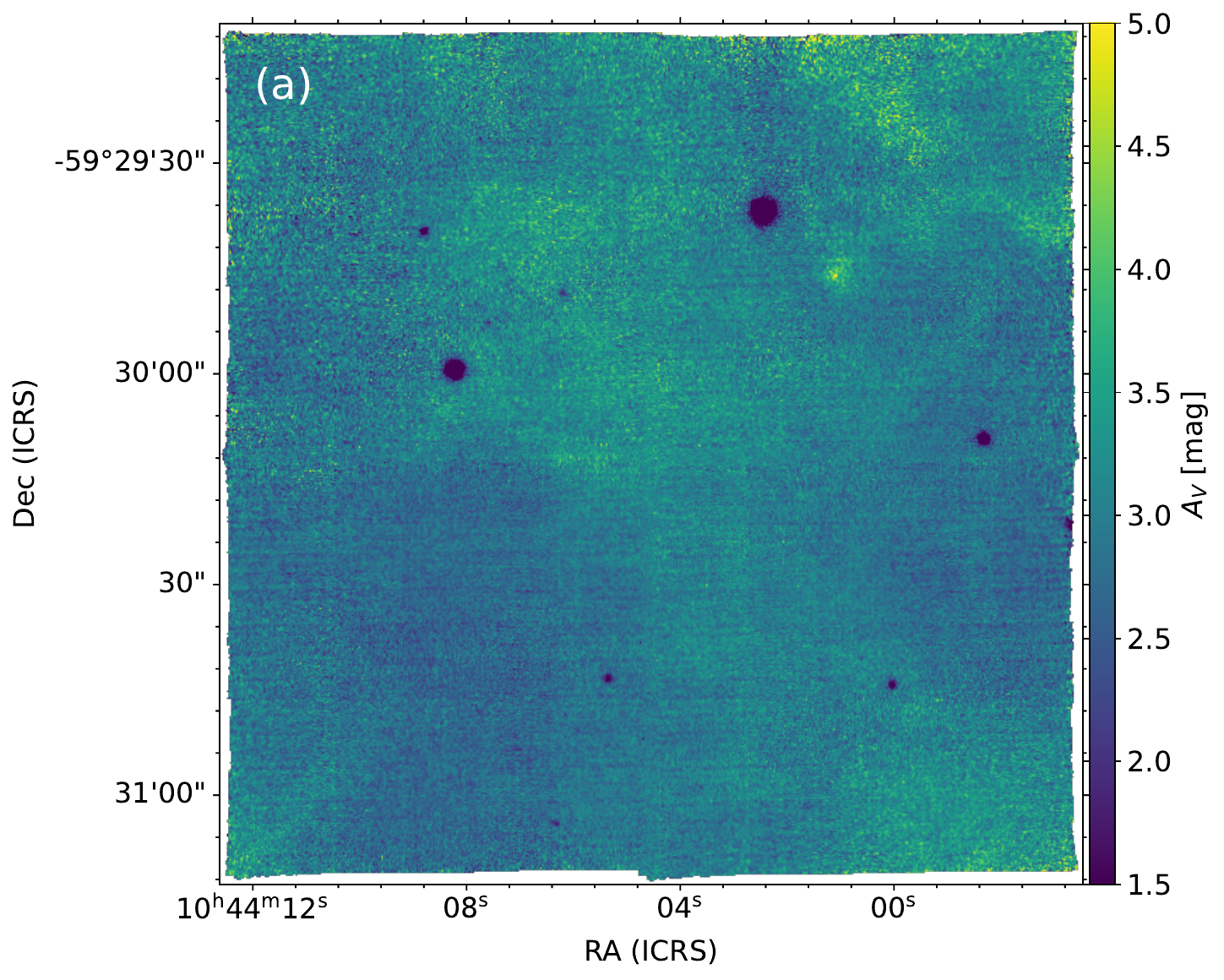}
    \includegraphics[width=\columnwidth]{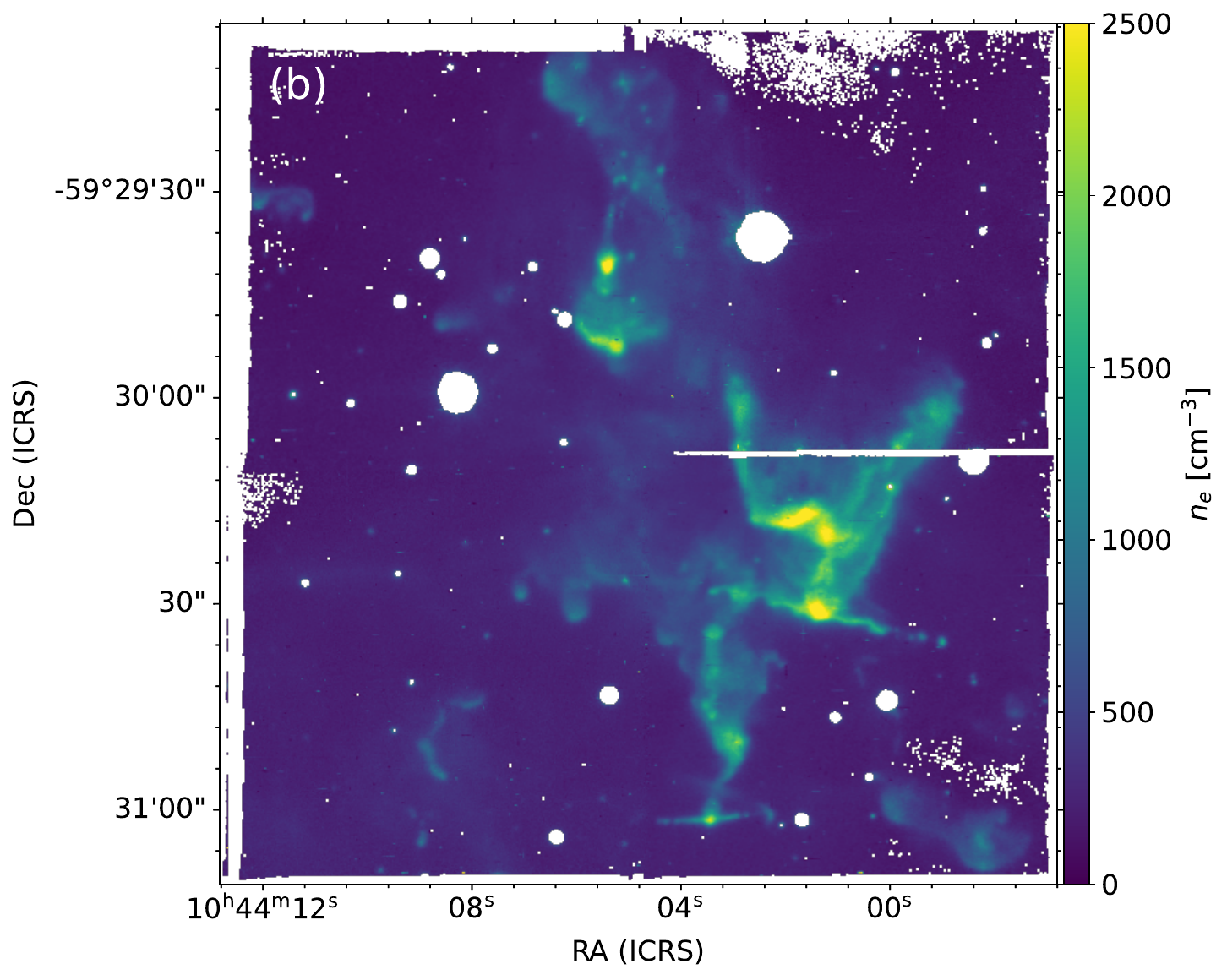}
    \includegraphics[width=\columnwidth]{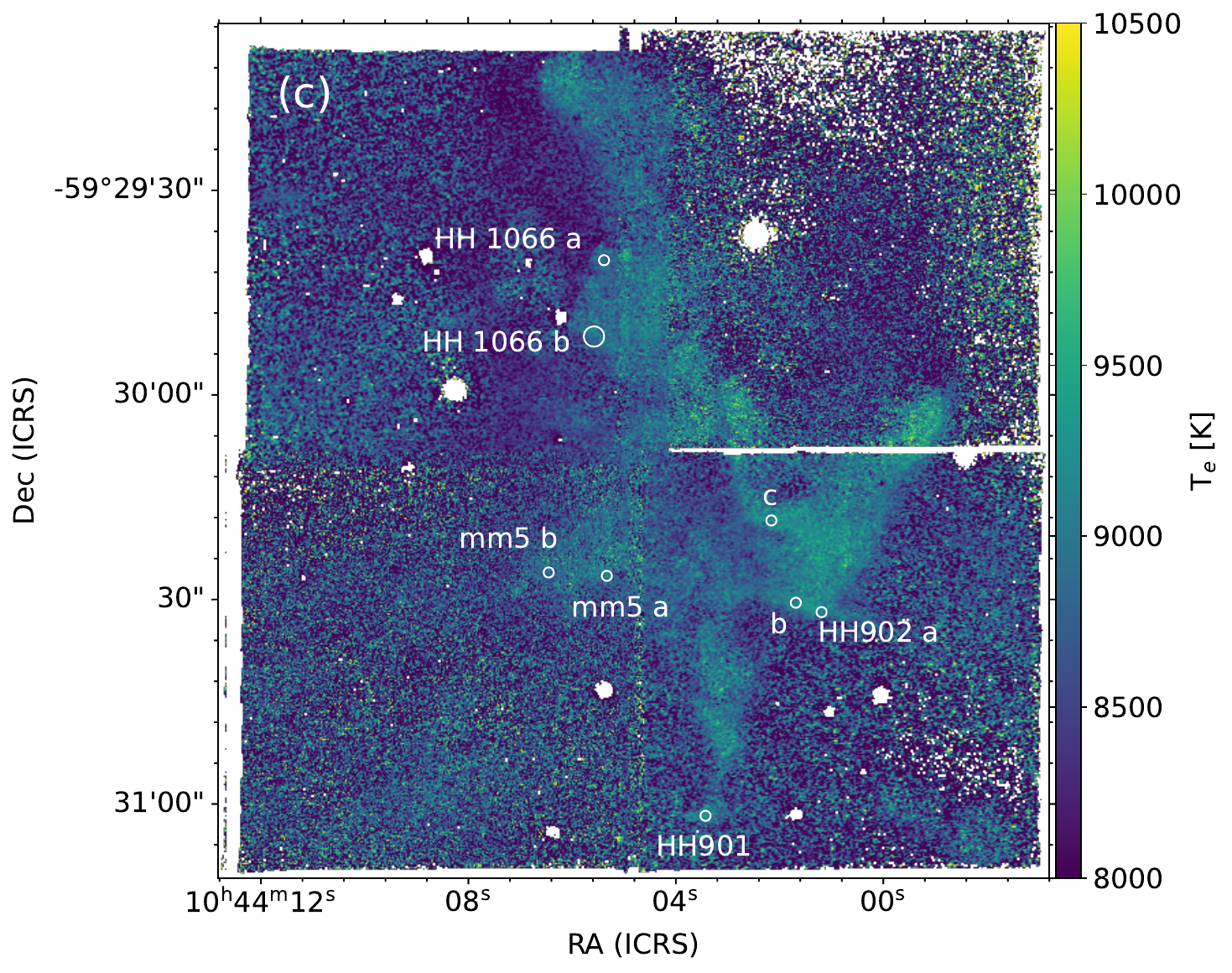}
    \includegraphics[width=\columnwidth]{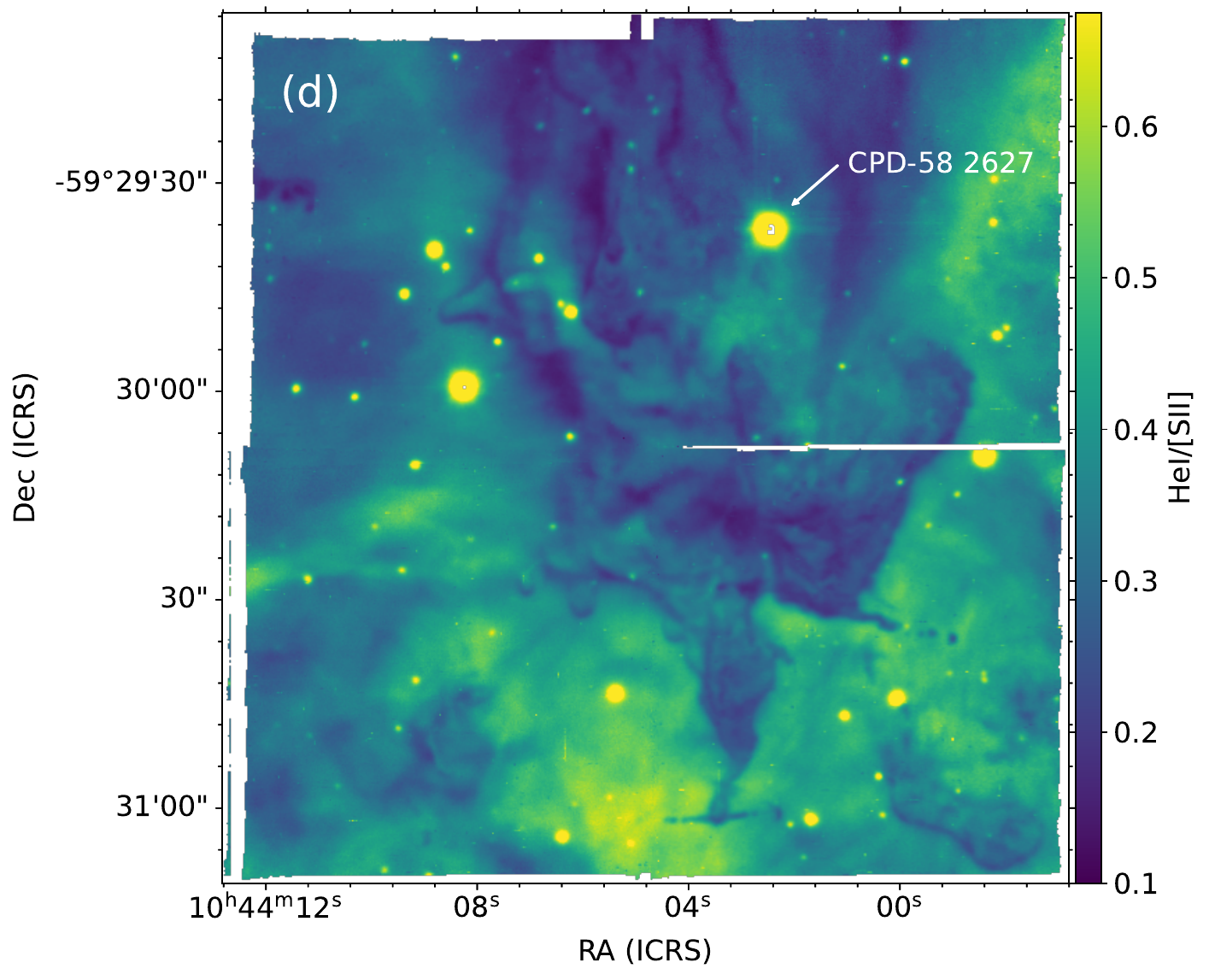}
        \caption{
        \textbf{(a)} $A_V$ map of the \mm, derived from the short exposures (see also Appendix~\ref{A:AV}). 
        \textbf{(b)} Electron density ($n_e$) derived from the ratio [S~{\sc ii}] $\lambda 6731 /6717$~\AA. 
        \textbf{(c)} Electron temperature ($T_e$) derived from [N~{\sc ii}] $\lambda 5755 / \lambda 6584 + \lambda 6548$~\AA. 
        \textbf{White circles show the regions used to compute the quantities in Table~\ref{t:props}.} 
        \textbf{(d)} Ionization map derived from the ratio He~{\sc i} $\lambda 6678$~\AA\ / [S~{\sc ii}] $\lambda 6717$~\AA. The O9.5V star CPD-58~2627, likely a foreground object (see Section~\ref{ss:excitation}), is labeled.  }
    \label{fig:ne_Te_maps}
\end{figure*}

\subsection{Extinction}\label{ss:extinction}

The line-of-sight extinction was computed as a function of position using {\sc PyNeb} \citep{luridiana2013,luridiana2015}. 
To compute the extinction toward the \mm, we use the \citet{ccm1989} extinction curves with $R_V = 4.4 \pm 0.2$ as measured toward Carina by \citet{hur2012}.
We assume Case~B recombination for gas with a temperature $T=10^4$~K and an electron density $n_e = 10^3$~cm$^{-3}$. 
Under these assumptions, the intrinsic flux ratio is (H$\alpha$/H$\beta$)$_{\mathrm{int}}$ = 2.86 \citep{osterbrock2006}. 

H$\alpha$ emission is saturated in the long exposures along the brightest ionization fronts (those that face Tr14). 
Fortunately, H$\alpha$ is not saturated in the short exposures. 
Extinction values derived from the short and long exposures are consistent to within $\sim 0.1$~mag everywhere except for where the H$\alpha$ emission is saturated in the long exposures. 
$A_V$ maps computed from both the long and short exposures and the difference between them are shown in Figure~\ref{fig:AV_maps} in Appendix~\ref{A:AV}. 
We use the extinction correction measured from the short exposures to correct the fluxes in both the short and long exposures. 
The corresponding $A_V$ map is shown in Figure~\ref{fig:ne_Te_maps}. 

The average extinction toward the \mm\ computed this way is $2.9 \pm 0.4$~mag (median $A_V=2.9$~mag). 
The $A_V$ varies smoothly over the observed region with slightly higher values in the northern regions of the mosaic. 
\citet{itrich2024} derived $A_V$ during spectral typing of the point sources in Tr14, including 91 sources from their high-confidence catalog in the region around the \mm. 
The average $A_V$ derived for these point sources is slightly lower, $1.9 \pm 0.9$~mag, with the highest individual source extinction of 4.5~mag. 
The high-confidence catalog from \citet{itrich2024} excludes sources with significant background contamination and therefore does not include many sources in or near the \mm. 
Stars removed due to high background contamination show a larger spread of $A_V$ values, with estimates as high as 7~mag.

\subsection{Electron temperature and density}\label{ss:Te_ne}

\begin{table*}
	\centering
	\caption{Properties at pillar heads. Columns are (1) the location where values are extracted; (2) the projected distance from Tr14; (3) radius of curvature at the pillar surface; (4) solid angle; (5) local ionizing photon luminosity; (6) electron density reported as the mean and standard deviation; (7) mean and standard deviation of the electron temperature; and (8) mass-loss rate due to photoevaporation. }
	\label{t:props}
	\begin{tabular}{lrcccccc} 
		\hline
		Location & d$_{\mathrm{proj}}$ & r$_{\mathrm{surf}}$ & $\Omega$ & log($Q_{0, \mathrm{local}}$) & $n_e$ & $T_e$ & $\dot{M}$ \\ 
         & [pc] & [pc] & [sr] & photons s$^{-1}$ & [cm$^{-3}$] & [K] & M$_{\odot}$~yr$^{-1}$ \\ 
		\hline
		HH901 & 1.3 & 0.01 & $3.5 \times 10^{-4}$ & 46.89 & $1825 \pm 390$ & $10392 \pm 253$ & $2.0 \pm 0.5 \times 10^{-7}$ \\ 
		
        HH902~a & 1.6 & 0.07 & $1.2 \times 10^{-2}$  & 48.42 & $2214 \pm 378$ & $10090 \pm 177$ & $1.2 \pm 0.3 \times 10^{-5}$ \\
        HH902~b & 1.6 & 0.07 & $1.2 \times 10^{-2}$ & 48.41 & $1908 \pm 155$ & $9990 \pm 125$ & $1.0 \pm 0.2 \times 10^{-5}$ \\
        HH902~c & 1.8 & 0.03 & $2.3 \times 10^{-3}$ & 47.70 & $2131 \pm 246$ & $9878 \pm 164$ & $2.7 \pm 0.5 \times 10^{-6}$ \\
		
        HH1066~a & 2.2 & 0.01 & $2.7 \times 10^{-4}$ & 46.76 & $2497 \pm 435$ & $10487 \pm 346$ & $5.8 \pm 1.3 \times 10^{-7}$ \\
        HH1066~b & 2.1 & 0.04 & $2.5 \times 10^{-3}$ & 47.74 & $1687 \pm 294$ & $10276 \pm 264$ & $3.4 \pm 0.8 \times 10^{-6}$ \\
        
        mm5~a & 1.8 & 0.05 & $5.5 \times 10^{-3}$ & 48.08 & $697 \pm 43$ & $10591 \pm 351$ & $2.1 \pm 0.3 \times 10^{-6}$ \\
        mm5~b & 1.8 & 0.03 & $1.9 \times 10^{-3}$ & 47.61 & $474 \pm 25$ & $10143 \pm 241$ & $5.1 \pm 0.8 \times 10^{-7}$ \\
        
		\hline
	\end{tabular}
\end{table*}
\begin{figure}
    \includegraphics[width=\columnwidth]{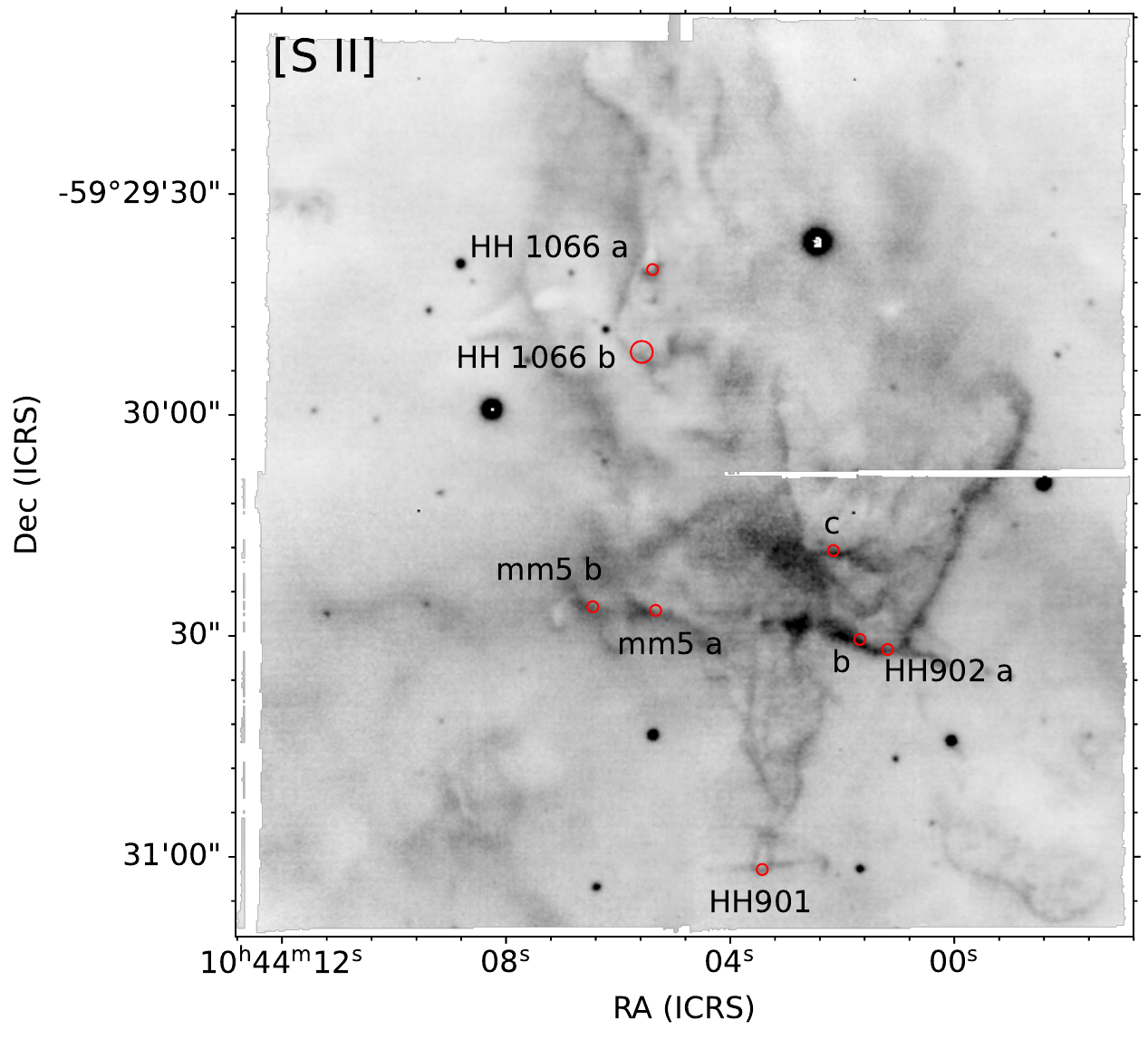}
    \caption{[S~{\sc ii}] $\lambda6717$\AA\ map with red circles showing where the quantities in Table~\ref{t:props} were computed (same as in Fig.~\ref{fig:ne_Te_maps}). }
    \label{fig:sii_map}
\end{figure}

We computed the electron temperature, $T_e$, and density, $n_e$, at every point in the map. 
To compute $T_e$, we use the ratio of lines with very different excitation energies, in this case [N~{\sc ii}] $\lambda 5755 / \lambda 6548,6583$ \citep[see, e.g.,][]{osterbrock2006,piembert2017}. 
The relative population of the levels provides a diagnostic of the plasma temperature. 
For the \mm, we obtain similar results if we use the ratio [S~{\sc iii}] $\lambda 6312 / \lambda 9068$ to compute the temperature. 
To compute $n_e$, we use the ratio of lines that originate from similar energy levels but have a large difference in collision strengths, in this case [S~{\sc ii}] $\lambda 6731 / \lambda 6716$. 
We use {\sc PyNeb} to compute these inter-related quantities simultaneously. 
Maps of the electron density and temperature are shown in Figure~\ref{fig:ne_Te_maps}. 
Uncertainties in these derived quantities depend on the signal-to-noise ratio of the lines, and thus vary over the map. In the following analysis, we focus on well-detected regions in and near the \mm, and thus determine uncertainties directly from the variations in each extracted region.

High electron densities, often in excess of 10$^3$~cm$^{-3}$, trace 
the heads of pillars and globules, especially near the driving sources of the famous HH jets. 
Equally high densities reveal compressed ridges without jet-driving YSOs behind them, such as the ionization front due south of HH~1066 and the ridge in the middle of the HH~902 pillar. 

Most of the highest density regions are perpendicular to the incident radiation from Tr14. 
One exception is a vertical spine of high densities that trace the length of the HH~901 pillar. 
High density clumps extend $\sim 28$\arcsec, reaching north from the head of the HH~901 pillar to the base of the HH~902 pillar.

The famous jets HH~901 and HH~902 also stand out as high density structures. 
Densities in the jets are $\gtrsim 10^3$~cm$^{-3}$, more than twice the density in the more diffuse gas in the pillars (away from the high-density edges) and nearly an order of magnitude higher than the ambient gas in the H~{\sc ii} region. 
The body of the smaller HH~1066 jet is difficult to resolve with these seeing-limited observations.
However, high densities at the head of the globule are elongated to the northwest, in the same direction as the jet axis seen in high resolution observations from the \emph{Hubble Space Telescope} \citep{reiter2013}. 
The HH~1066 bow shock can be seen as an additional spot of high density $\sim 2^{\prime\prime}$ to northwest of the pillar.  

A temperature map shows much less structure with hot gas throughout the imaged area. 
Electron temperatures along the ionization front are typically near 10,000~K (see Figure~\ref{fig:ne_Te_maps}). 
We report values for the electron densities and temperatures at a few locations near the pillar tips (shown in Figure~\ref{fig:sii_map}) in Table~\ref{t:props}.

\subsection{Excitation}\label{ss:excitation}

We consider the ratio of lines with similar ionization potentials, 
He~{\sc i} $\lambda 6678$~\AA\ / [S~{\sc ii}] $\lambda 6717,6731$~\AA, to estimate the degree of ionization in the gas. 
Ratios $\gtrsim 1$ are found in highly ionized H~{\sc ii} regions like M17 \citep{glushkov1998}. 
As shown in Figure~\ref{fig:ne_Te_maps}d, the most highly excited gas in the \mm\ is found to the south and the west. 
The degree of ionization drops sharply at the pillar boundaries where densities are high (e.g., the ridge in the center of the HH~902 pillar). 
Some higher ionization gas can be seen in the northwest of the \mm, near the O9.5V star CPD-58 2627. 
The emission is diffuse and does not trace the structure of the \mm, hinting that CPD-58 2627 and the associated excited gas may be in the foreground of the \mm. 
Unfortunately, uncertainties on the \emph{Gaia} parallax of CPD-58 2627 are large, so we cannot confirm this hypothesis \citep{goppl2022,gaia_dr3_2023}.

The HH~901 and HH~902 jets stand out as low-ionization structures in the He~{\sc i} $\lambda 6678$~\AA\ / [S~{\sc ii}] $\lambda 6717,6731$~\AA\ ratio map despite their high intensity in hydrogen recombination lines like H$\alpha$. 
Low excitation features in both jets include multiple knots beyond the continuous inner jet and the terminal bow shocks seen to the west of the \mm.

\begin{figure}
	\includegraphics[width=\columnwidth]{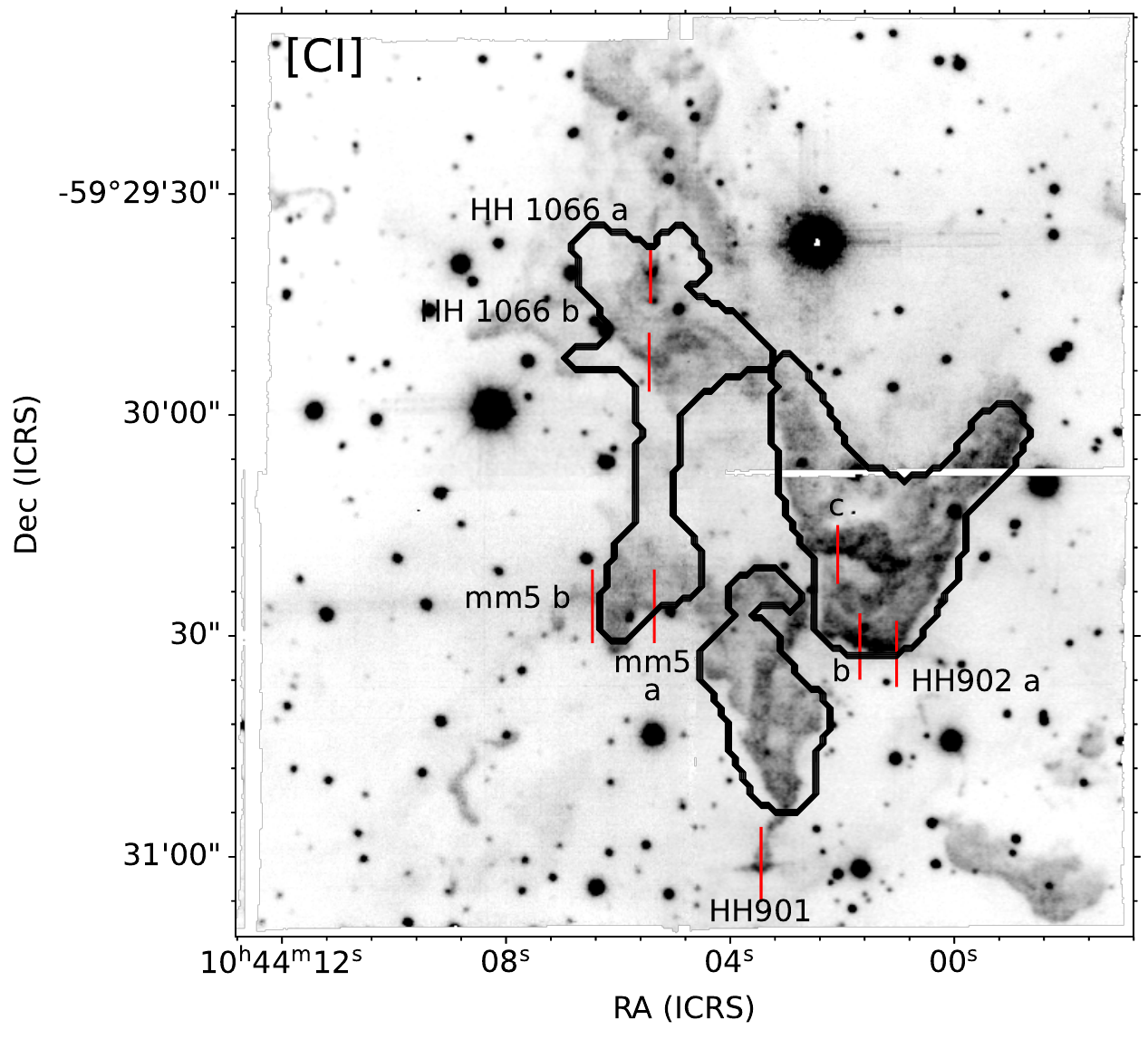}
    \caption{\textbf{Left:} Map of [C~{\sc i}] 8727~\AA\ showing the entire \mm. Red lines show the location where the intensity tracings in Figure~\ref{fig:ifronts} are taken. Single black contours show the sub-pillars (the HH~901 pillar, etc) definitions from \citet{reiter2023_MM}. }
    \label{fig:ci_map}
\end{figure}

Emission from the neutral carbon line, [C~{\sc i}] $\lambda 8727$\AA, is 
wide-spread in the \mm\ (see Figure~\ref{fig:ci_map}). 
The ionization potential of C is 11.26~eV, so this line is often found coincident with dissociating H$_2$ in photodissociation regions \citep[PDRs;][]{escalante1991,esteban2004}, including photoevaporating protostellar jets \citep[e.g.,][]{reiter2019_tadpole}, and photoevaporating protoplanetary disks \citep[proplyds; e.g.,][]{haworth2023,goicoechea2024}. 
In the \mm, bright [C~{\sc i}] emission primarily traces regions with high $n_e$ along the pillar edges. 
This includes [C~{\sc i}] emission that extends east-west from the head of the HH~901 pillar. 
The morphology resembles the extended H$_2$ emission seen by \citet{hartigan2015}, suggesting that both trace the jet/outflow. 
Extended [C~{\sc i}] emission cannot be readily identified for HH~902 or HH~1066.

\begin{figure*}
    \includegraphics[width=\columnwidth]{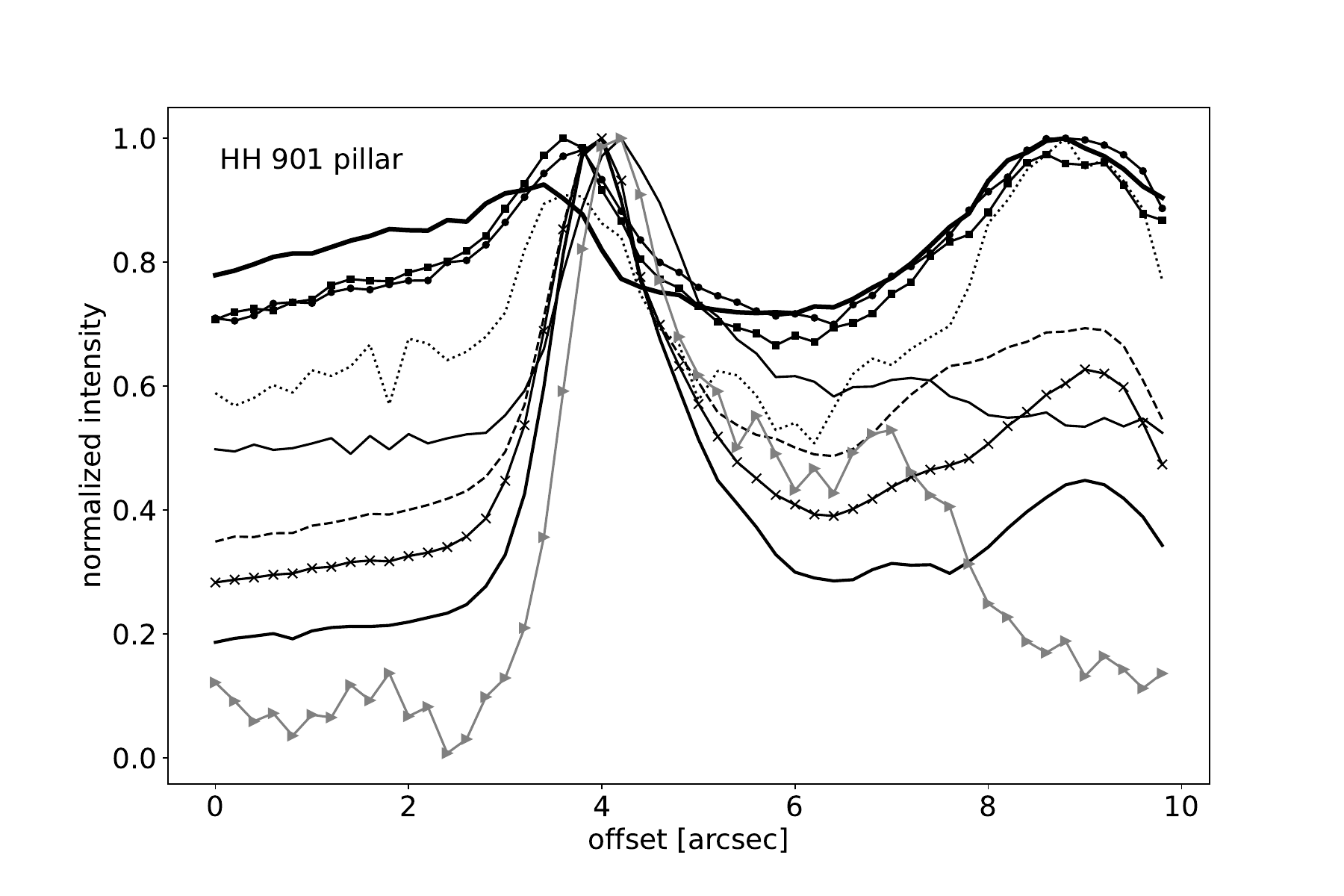}
    \includegraphics[width=\columnwidth]{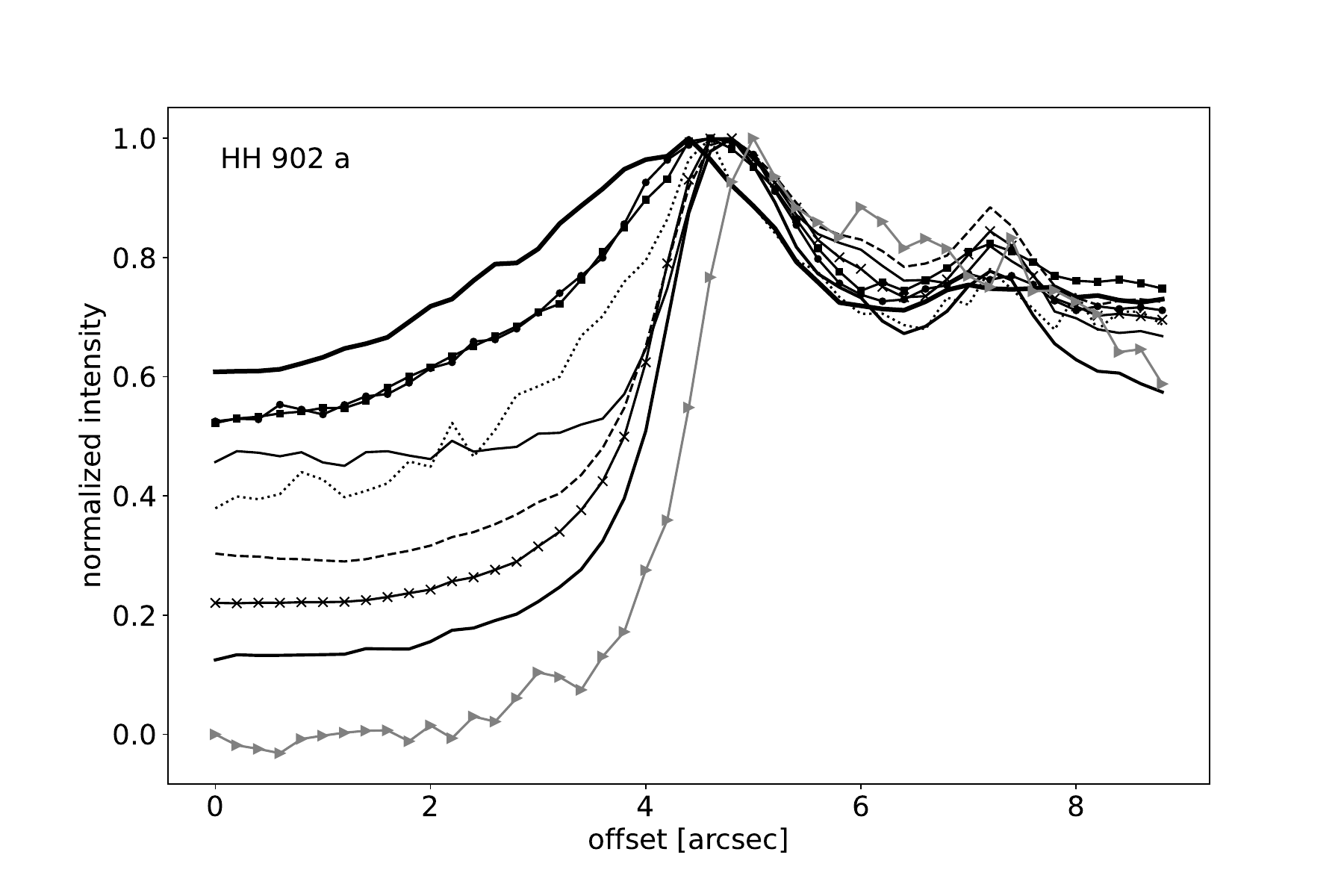}
    \includegraphics[width=\columnwidth]{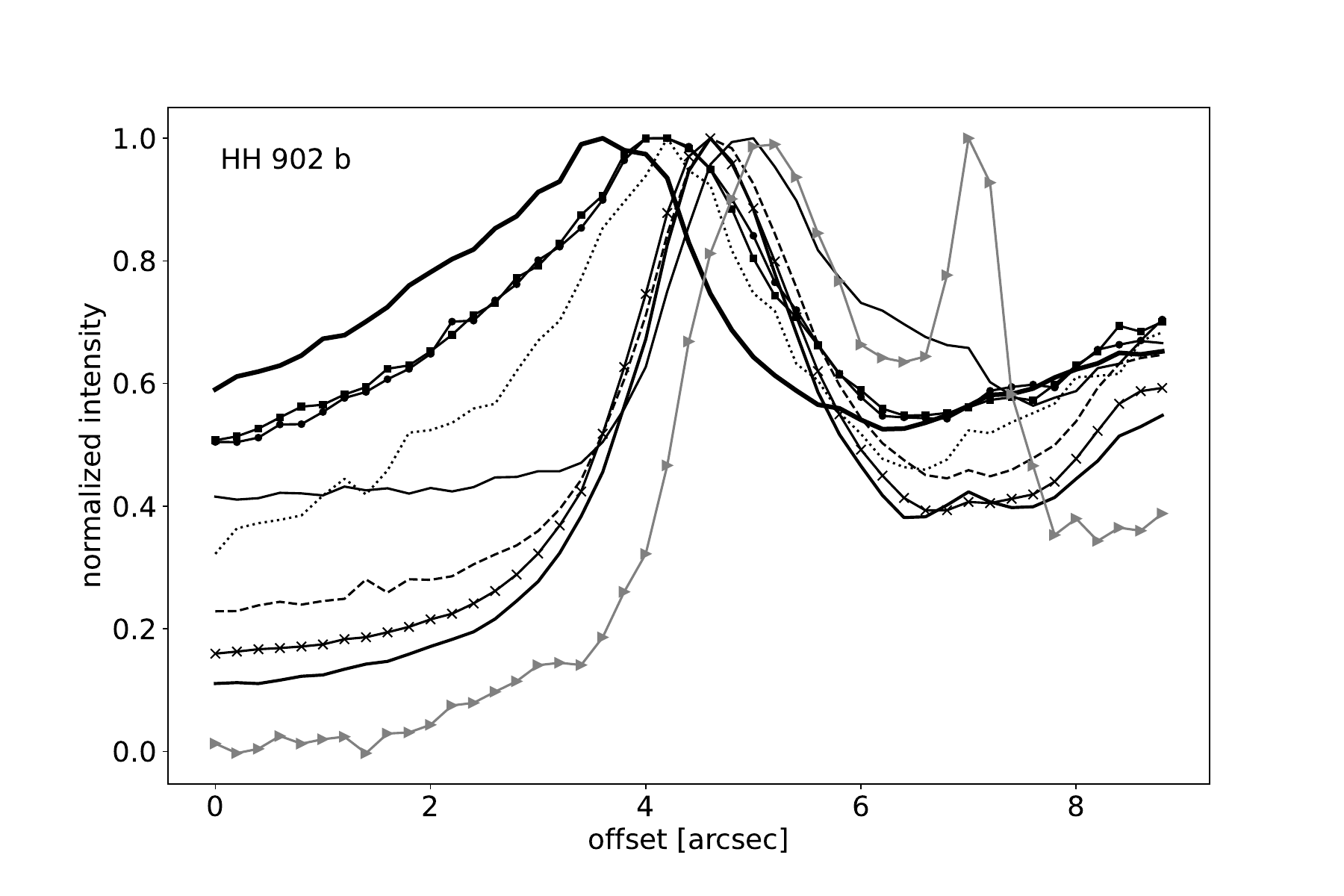}
    \includegraphics[width=\columnwidth]{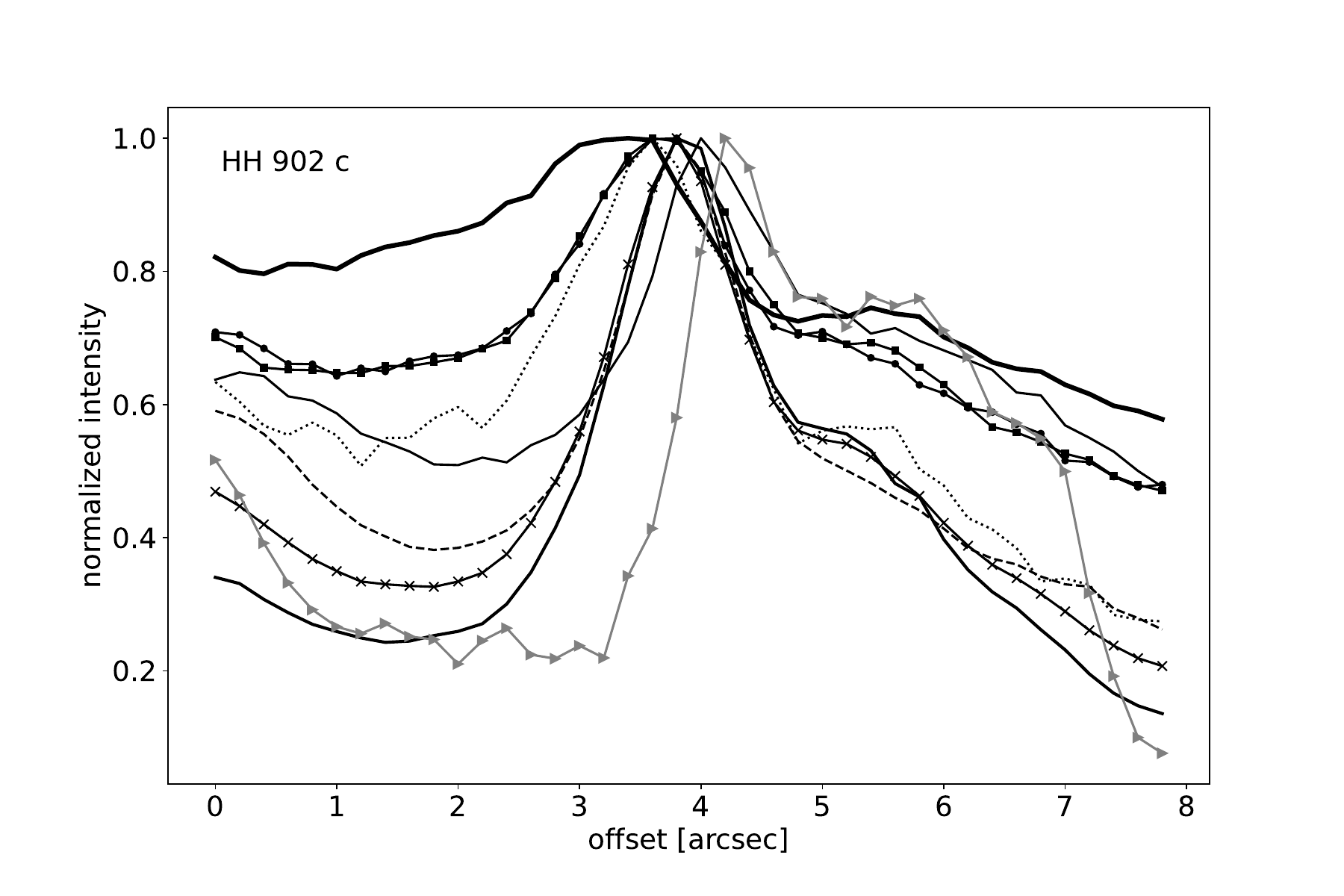}
    \includegraphics[width=\columnwidth]{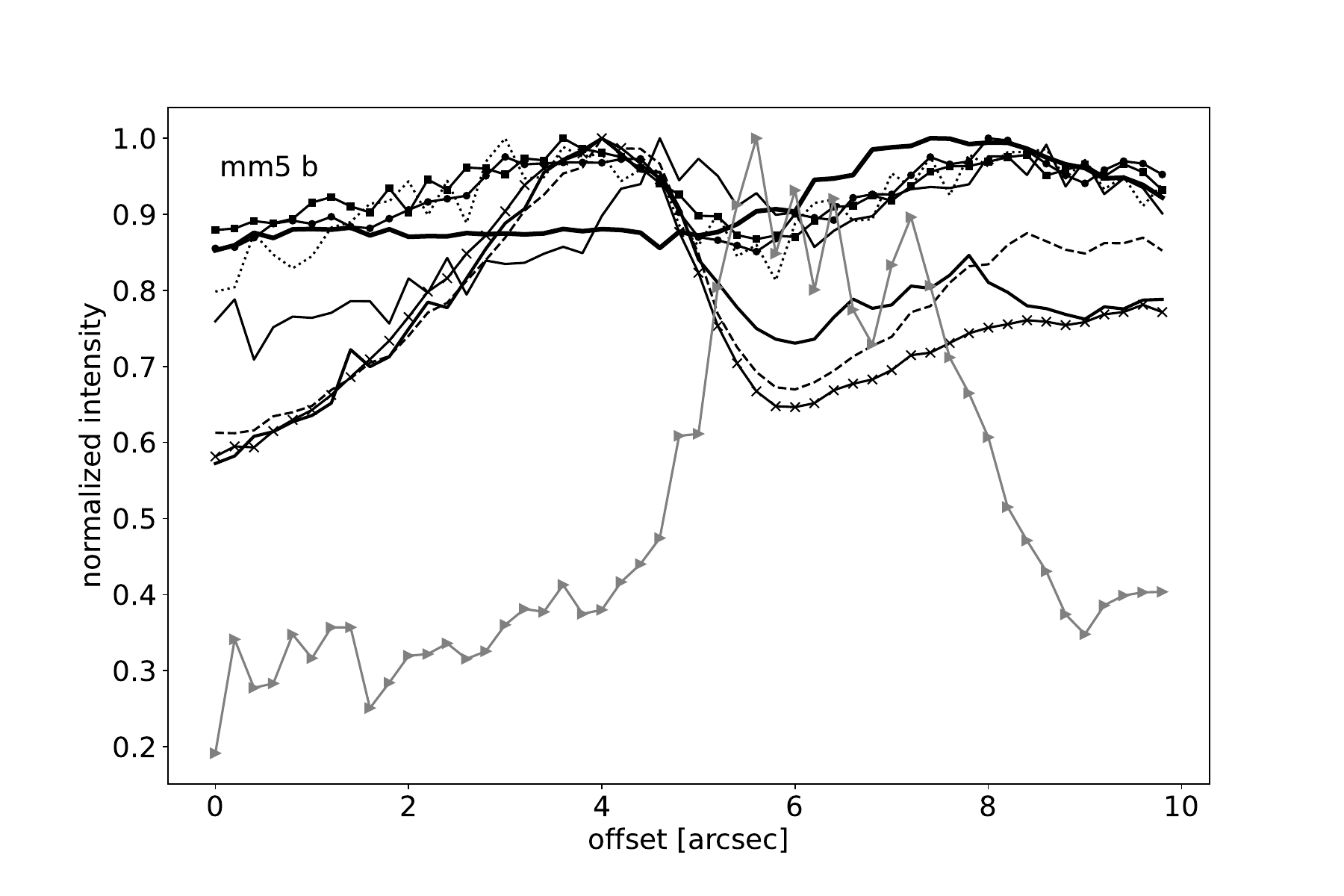}
    \includegraphics[width=\columnwidth]{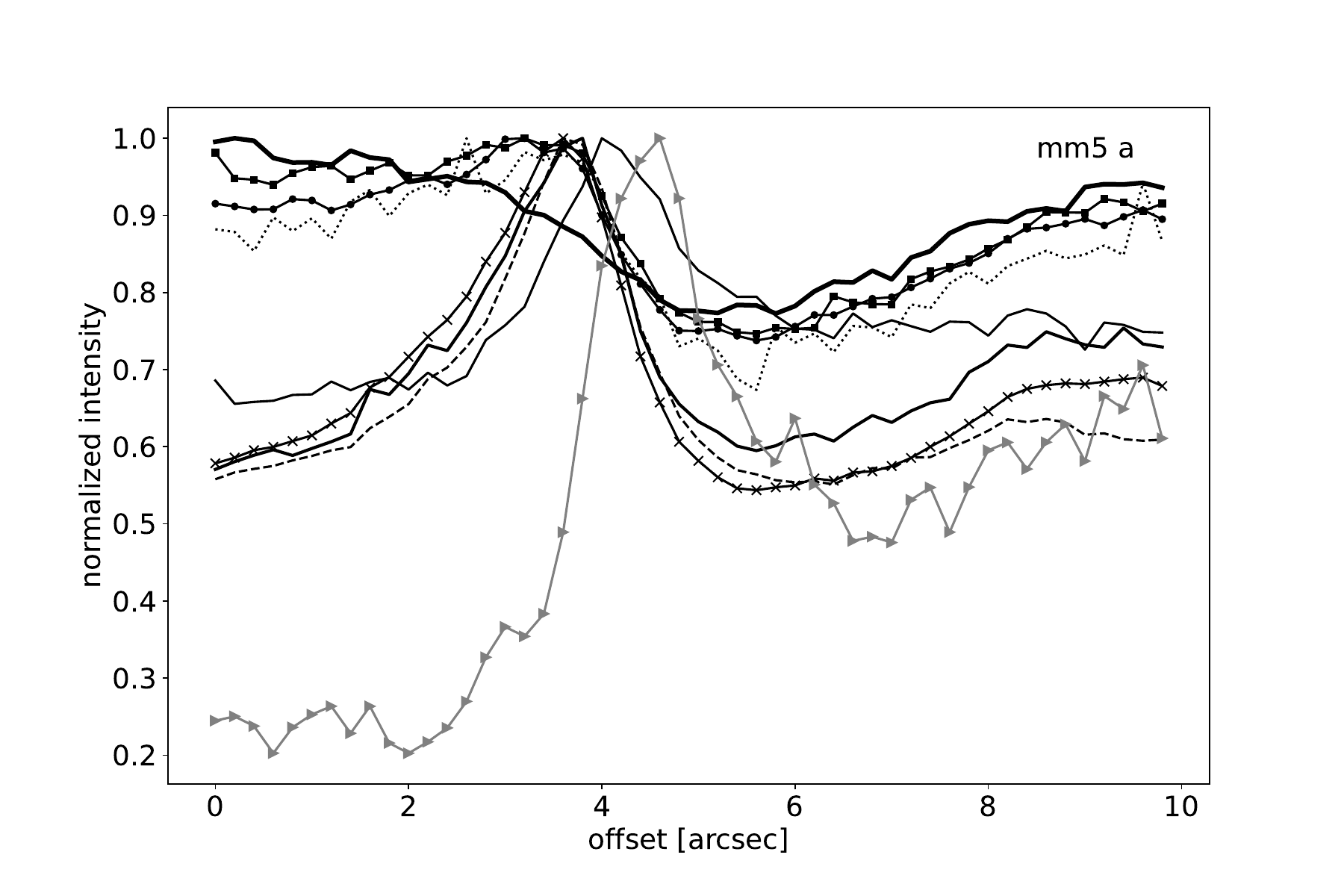}
    \includegraphics[width=\columnwidth]{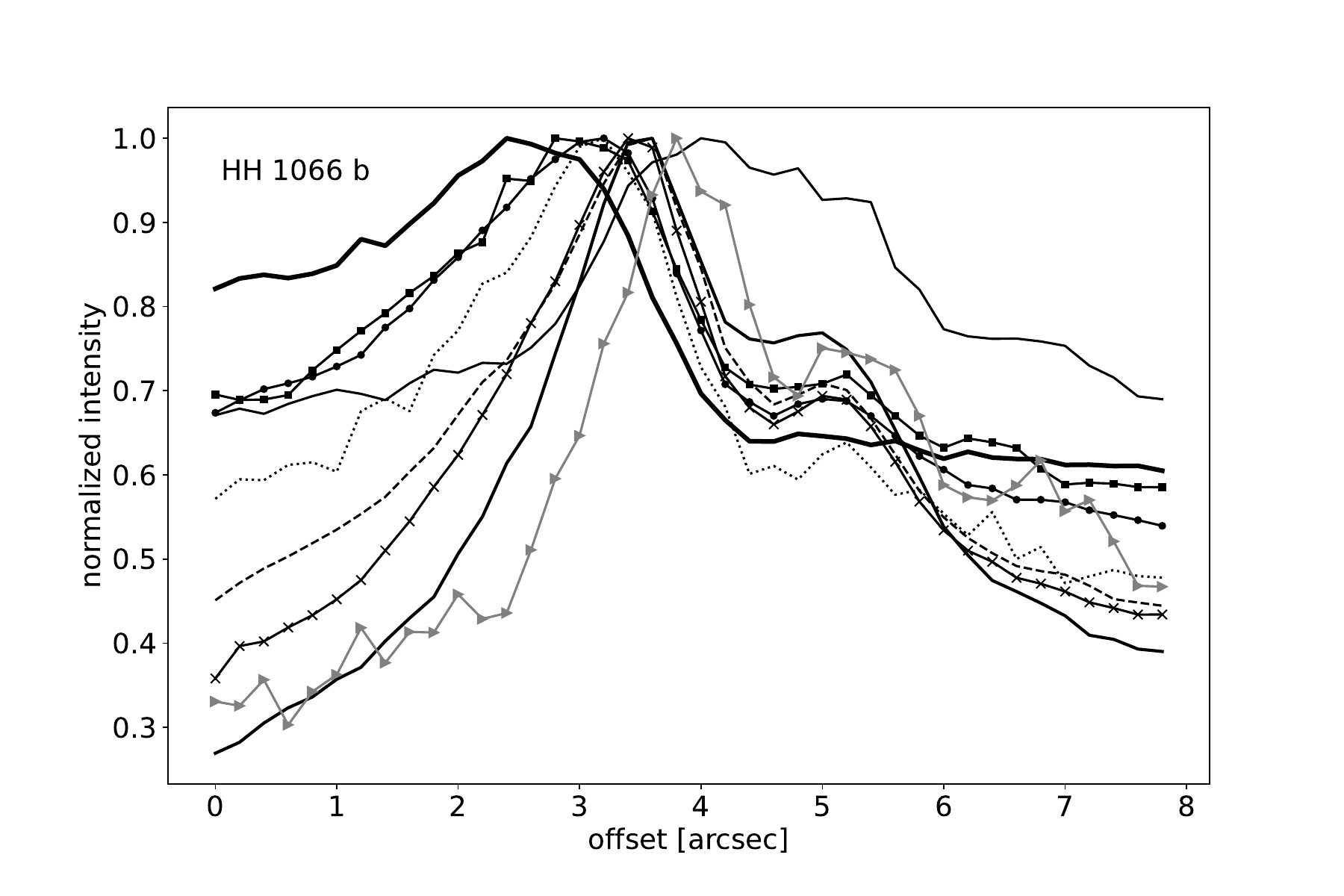}
    \includegraphics[width=\columnwidth]{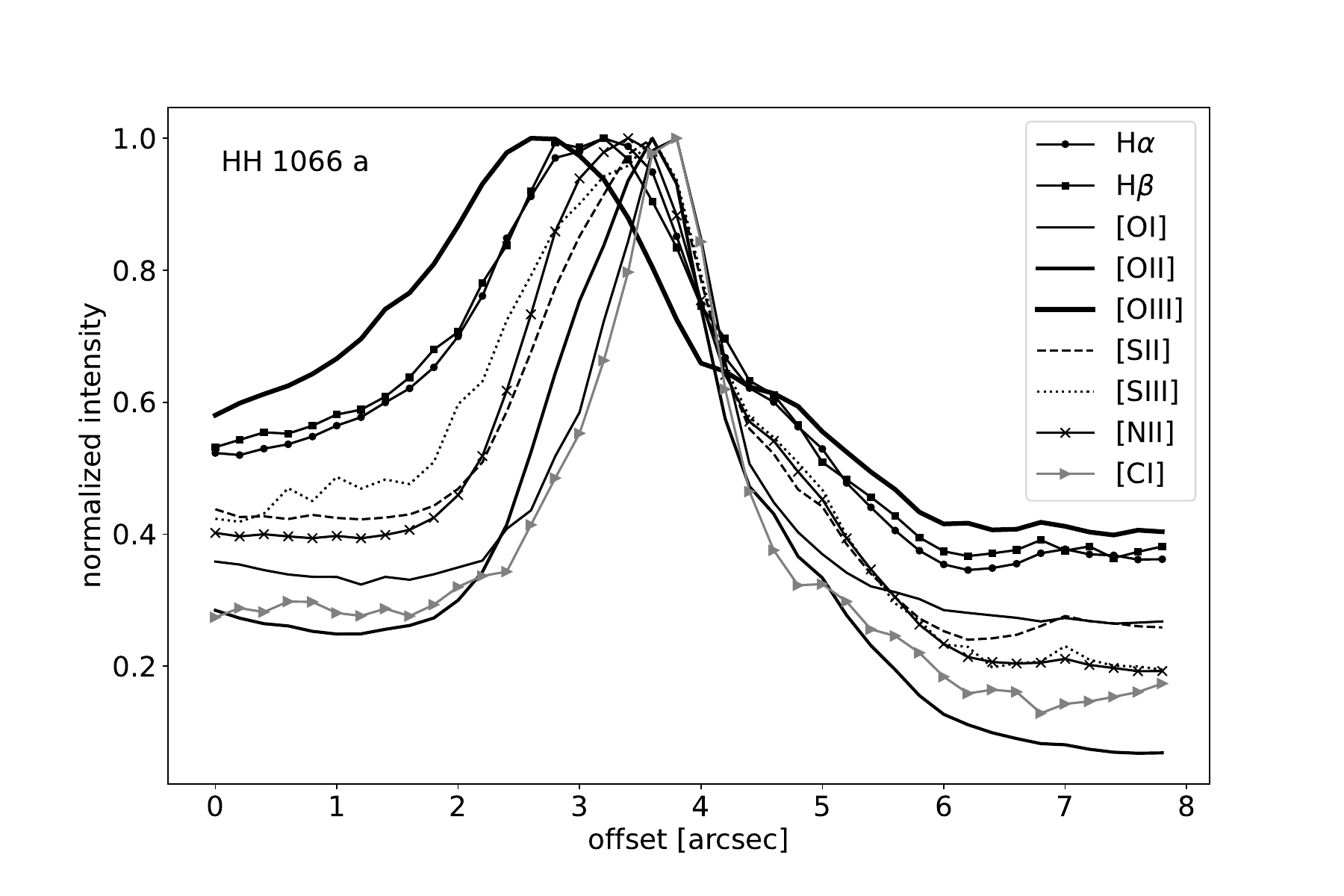}
    \caption{Locations of tracings shown in Figure~\ref{fig:ci_map}. 
    } 
    \label{fig:ifronts}
\end{figure*}

Finally, we extract the intensity profiles of emission lines of different excitation to trace the structure of the ionization fronts.  
Red lines in Figure~\ref{fig:ci_map} show the locations where we extract intensity profiles. 
We target ionized interfaces that coincide with the millimeter continuum sources identified by \citet{reiter2023_MM}. 
We also sample multiple locations along large ionization fronts like the HH~902 pillar. 
The intensity tracings are plotted in Figure~\ref{fig:ifronts}.

Intensity tracings at pillar tips ahead of jet-driving YSOs (HH~901~mm, HH~902~mm, HH~1066~mm) have the smallest offsets between the peaks of different emission lines. 
Offsets are somewhat larger in other high density interfaces like HH~902~c and HH~1066~b. 
All of these regions have high $n_e$ ($>$10$^3$~cm$^{-3}$ see Table~\ref{t:props}). 
Each of these highly irradiated interfaces is associated with a bright 1.3~mm continuum source; several also have emission from high density lines like C$^{18}$O and $^{13}$CS that indicate the presence of a deeply embedded core \citep{reiter2023_MM}. 
Offsets between intensity peaks are largest in ionized interfaces that do not have any evidence for an embedded YSO immediately behind the ionization front (e.g., HH~902~b). 


\subsection{Cores, clumps and young stellar objects}\label{ss:ysos}

\begin{figure}
	\includegraphics[width=\columnwidth]{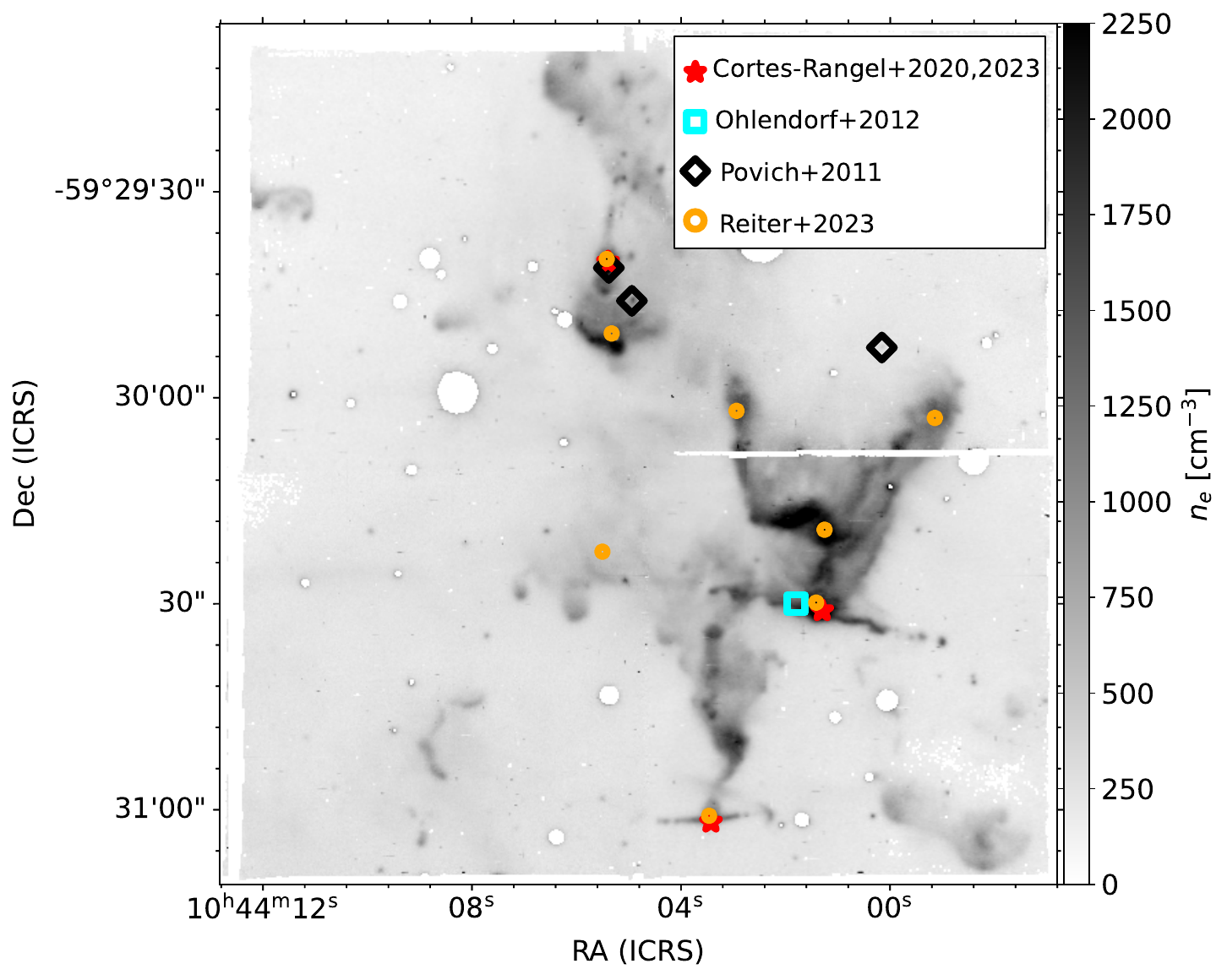}
	    \caption{A density map of the \mm\ overplotted with the YSOs identified from the mid-IR SEDs from \citet{povich2011} (black diamonds) and \citet{ohl12} (cyan squares). Millimeter continuum sources from \citet{cortes-rangel2020,cortes-rangel2023} are shown as red stars and those from \citet{reiter2024} as orange circles. 
        }
    \label{fig:density}
\end{figure}

The famous jets HH~901, HH~902, and HH~1066 were one of the first unambiguous signs of star formation in the \mm. 
Embedded young stellar objects (YSOs) have been harder to detect due to the large distance to the region (2.35~kpc). 
This is particularly acute at longer wavelengths where deeply embedded YSOs emit the majority of their emission. 
Nevertheless, a few previous works have identified candidate YSOs in the region. 
\citet{povich2011} produced the Pan-Carina YSO Catalog (PCYC) of intermediate-mass ($\sim 2-8$~\Msun) YSOs 
with masses and luminosities determined from model fits to the infrared spectral energy distribution (SED). 
These are shown as black diamonds in Figure~\ref{fig:density}. 
Three of the PCYC candidate YSOs reside in the \mm. 
\citet{reiter2016} identified PCYC~429 as the HH~1066 driving source. 
The other two YSOs lie in the northern part of the \mm, not clearly associated with any of the high-density gas identified with MUSE or ALMA. 
\citet{ohl12} identified an additional source at the head of the HH~902 pillar that has strong infrared emission detected with \emph{Herschel} (shown as a cyan square in Figure~\ref{fig:density}).

High-resolution ALMA observations have provided the best view of the embedded jet-driving YSOs \citep{cortes-rangel2020,cortes-rangel2023}.  
Disk+envelope masses of $\sim 0.1$~\Msun\ suggest that the jet-driving sources are low-mass stars (red stars in Figure~\ref{fig:density} show the sources observed at high resolution with ALMA). 
Marginally resolved observations of the disk around the HH~1066 driving source show that it is similar to disks seen around low-mass stars in local, more quiescent regions \citep[$r \sim 60$~AU; $M_{\mathrm{disk}} \gtrsim 45$~M$_{\mathrm{Jup}}$, see][]{mesa-delgado2016}. 
Larger scale but lower resolution ($\sim6$\arcsec) 1.3~mm observations from \citet{reiter2023_MM} revealed an additional 5 continuum sources in the \mm\ (shown as orange circles in Figure~\ref{fig:density}).

\citet{itrich2024} provided spectral types of hundreds of low-mass stars in Tr14, including several around the \mm. 
None of the stars in their catalog correspond to the deeply embedded jet-driving sources. 
Indeed, none of the candidate YSOs associated with the \mm\ are optically revealed, preventing accurate spectral typing. 
Some continuum emisision at the reddest wavelengths can be seen from the HH~1066 driving source (see the vertical emission in Figure~\ref{fig:hh1066_jet_vels}) but none from the HH~901~YSO or the HH~902~YSO. 

\subsection{Jets and outflows}\label{ss:jets}

\begin{figure}
	\includegraphics[width=\columnwidth]{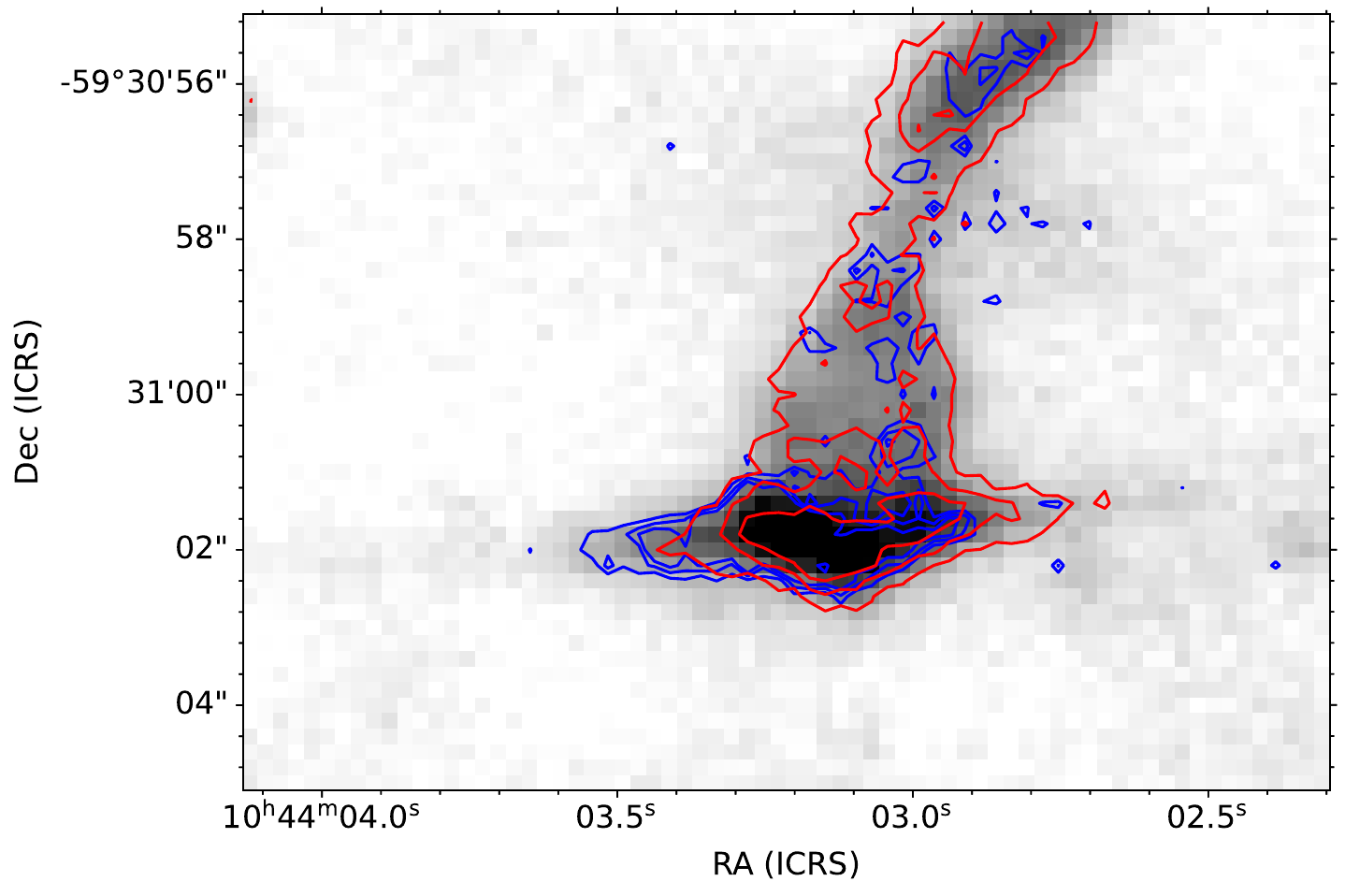}
    \caption{Map of [C~{\sc i}] 8727~\AA\ emission from HH~901. Blue contours show the emission as 8725.2~\AA\ and red contours show emission at 8728.9~\AA. Extended [C~{\sc i}] emission suggests that HH~901 is a photoevaporating molecular outflow. }
    \label{fig:hh901_ci}
\end{figure}

\begin{figure}
	\includegraphics[width=\columnwidth]{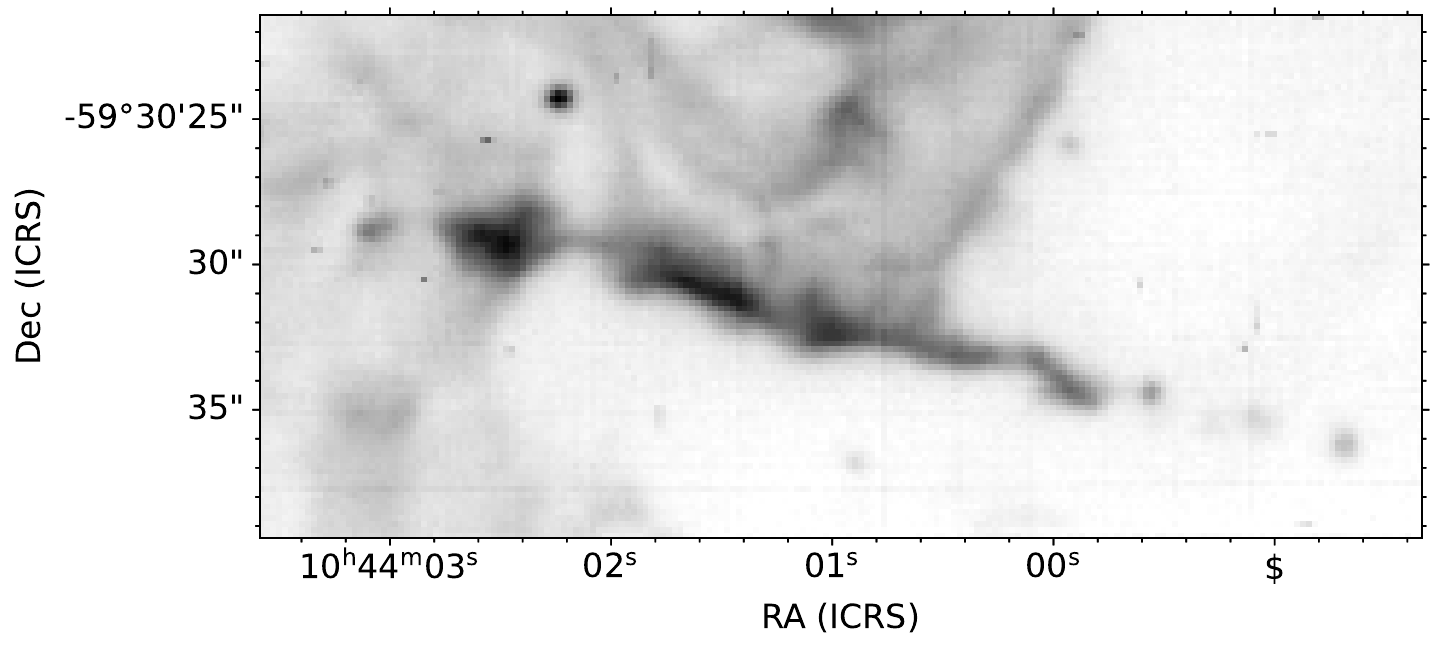}
    \caption{[O~{\sc i}] image of HH~902 showing the large amount of neutral material in the jet. }
    \label{fig:hh902_oi}
\end{figure}

\begin{figure}
	\includegraphics[width=\columnwidth]{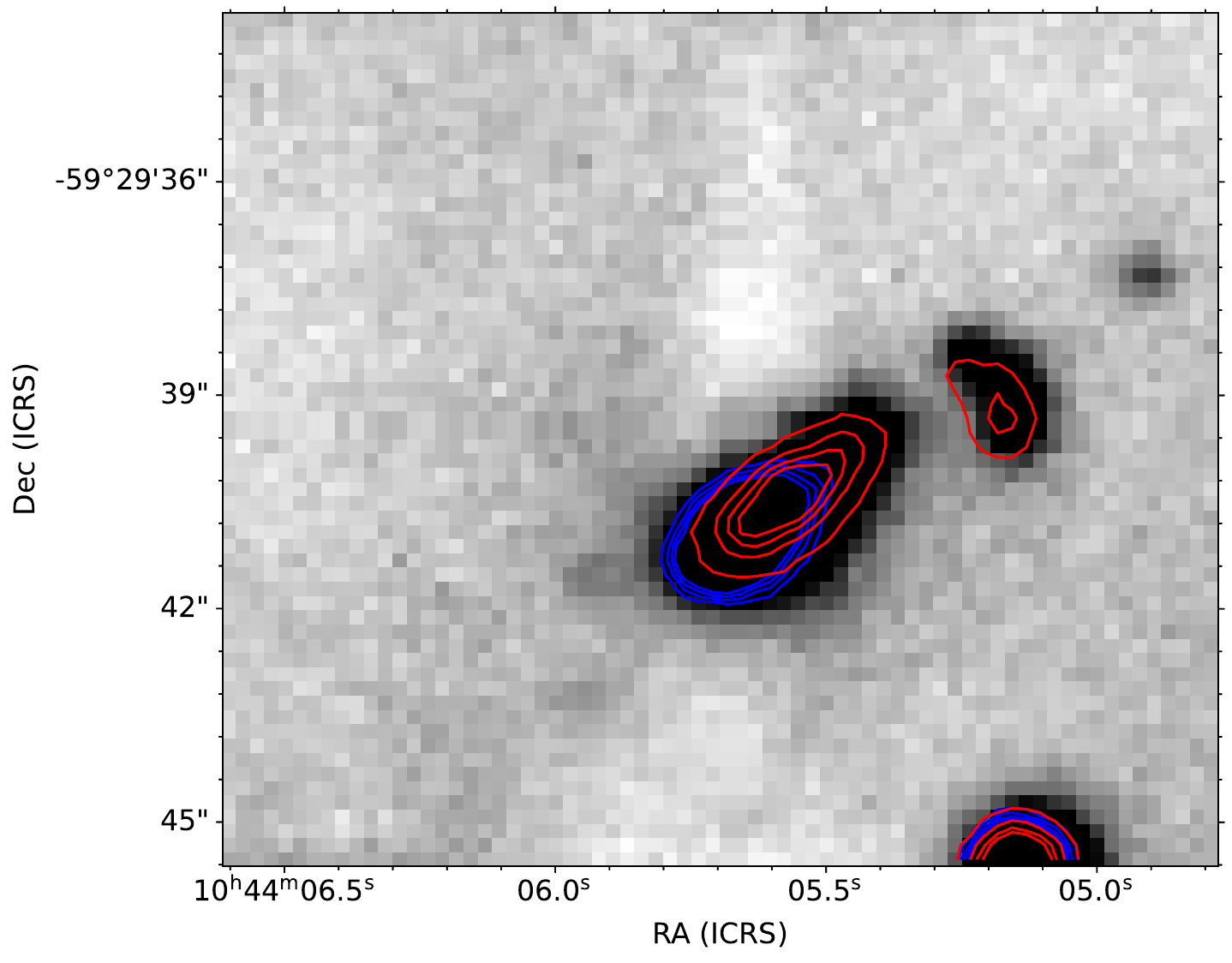}
    \caption{[Fe~{\sc ii}] 8617~\AA\ image of HH~1066 with blue contours showing emission in the range 8611.4--8615.2~\AA\ and red contours showing emission in the range 8617.7--8622.7~\AA.  }
    \label{fig:hh1066_feii}
\end{figure}

The \mm\ contain the famous Herbig-Haro (HH) jets \citep{smith2010,reiter2013,cortes-rangel2020}: HH~901 (see Figure~\ref{fig:hh901_ci}), HH~902 (see Figure~\ref{fig:hh902_oi}), and HH~1066 (see Figure~\ref{fig:hh1066_feii}). 
Electron densities derived from the ratio of [S~{\sc ii}] lines is high in all of the spatially resolved jets (see Section~\ref{ss:Te_ne}). 
Values of $n_e$ in HH~901 and HH~902 range from a few hundred to nearly 2000~cm$^{-3}$, consistent with earlier $n_e$ estimates from the H$\alpha$ emission measure \citep{smith2010}. 

\begin{figure}
	\includegraphics[width=\columnwidth]{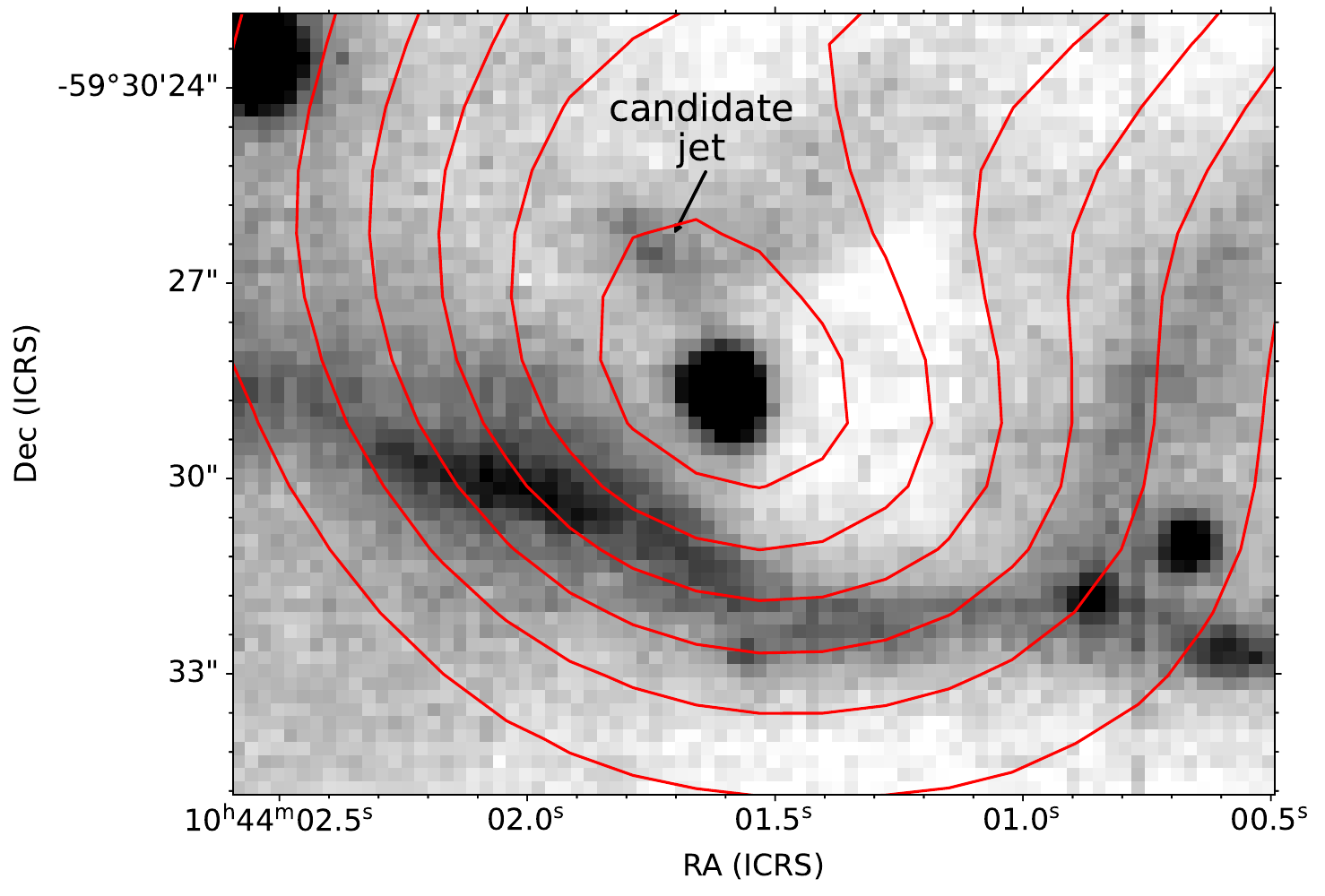}
    \caption{[Fe~{\sc ii}] 8617~\AA\ image showing candidate jet with C$^{18}$O contours overplotted. Peak of C$^{18}$O emission is slightly offset from the position of the star that appears to drive the jet.  }
    \label{fig:candidate_jet}
\end{figure}

We identify a new candidate jet located $\sim2$\arcsec\ north of HH~902 (see Figure~\ref{fig:candidate_jet}). 
Collimated emission is prominent in multiple [Fe~{\sc ii}] lines (and can also be seen in near-IR narrowband [Fe~{\sc ii}] imaging from \emph{HST}; \citealt{reiter2013,reiter2016}). 
The emission appears to originate from a nearby, IR-bright star. 
The candidate jet and driving source are coincident with the peak of C$^{18}$O emission seen at low resolution \citep[red contours on Figure~\ref{fig:candidate_jet};][]{reiter2023_MM}. 
\citet{itrich2024} estimated a spectral type of M3.0 and an $A_V=3.8$~mag for this source but note a large uncertainty on these parameters because of significant nebular contamination. 
Neither \citet{itrich2024} nor \citet{povich2011} identify an IR-excess associated with this source.

We show the velocity profiles of each jet in 
H$\alpha$, [S~{\sc ii}] $\lambda 6717$\AA, [O~{\sc i}] $\lambda6300$\AA, [C~{\sc i}] $\lambda8727$\AA, and [Fe~{\sc ii}] $\lambda8616$\AA\ using {\sc pvextractor}; these are shown in Figures~\ref{fig:hh901_jet_vels}-\ref{fig:cnj_jet_vels} in Appendix~\ref{A:jet_PVs}. 
To make the position-velocity (P-V) diagrams, we extract emission in a 1\arcsec\ wide region centered on the outflow. 
This complement of lines traces gas in different states, from highly ionized (H$\alpha$) to low ionization especially in regions that are shielded from the Lyman continuum ([S~{\sc ii}] and [Fe~{\sc ii}]) to neutral gas ([O~{\sc i}] and [C~{\sc i}]). 
Where H$\alpha$ emission is saturated in the long exposures (throughout HH~901 and HH~902), we use the short exposures to extract the H$\alpha$ velocities. 
For the H$\alpha$ emission from HH~1066 and all other emission lines from all jets, we use the long exposures.

All of the jets are spatially extended, suggesting that they lie no more than $\sim$45$^{\circ}$ away from the plane of the sky.  
For small tilt angles, we expect to see much of the outward velocity of the jets as proper motions \citep[see][]{reiter2014,reiter2017}. 
Despite the low velocity resolution of MUSE ($\gtrsim 100$~\kms\ at these wavelengths), the P-V diagrams reveal blueshifted emission in the eastern jet limb and redshifted to the west in both HH~901 and HH~902 (see Appendix~\ref{A:jet_PVs}).  
The clearest velocity difference is seen in [Fe~{\sc ii}] from HH~1066. 
As with HH~901 and HH~902, the eastern limb is blueshifted. 
Unusually, in HH~1066 [O~{\sc i}] traces a similar velocity profile to [Fe~{\sc ii}]. 
In other externally irradiated jets in Carina observed with MUSE, the [O~{\sc i}] morphology and kinematics more closely resemble the slower and wider-angle emission traced by H$\alpha$ than the fast, collimated jet seen in [Fe~{\sc ii}] \citep{reiter2019_tadpole}.

Only HH~901 shows extended [C~{\sc i}] emission (see Figure~\ref{fig:hh901_ci}). 
The velocity structure is marginally resolved but 
consistent with the jet orientation seen in other lines (see Figure~\ref{fig:hh901_jet_vels}). 
This line tends to be seen in the same region as dissociating molecular gas, including externally irradiated molecular outflows \citep[see][]{reiter2024}. 
This may be the irradiated counterpart to the molecular outflow seen inside the globule by \citet{cortes-rangel2020}.

We do not detect red or blueshifted emission from the [Fe~{\sc ii}] P-V diagram of the candidate jet (the candidate jet is not detected in the other lines). 
As with the other outflows in the \mm, this candidate jet may lay close to the plane of the sky and therefore have small radial velocities. 
In the future, proper motions measurements may confirm the outflow nature of this object.

\section{Discussion}\label{s:discussion}

Several papers have examined the structure and kinematics in the molecular gas in different portions of the Carina Nebula \citep{klaassen2020,rebolledo2020,menon2021,hartigan2022}. 
However, few of these studies overlap with measurements of the ionized gas properties \citep[e.g.,][]{mcleod2016}, preventing the direct comparison between the input from the nearby high-mass stars and the resulting structure in the cold, molecular gas. 
Making the explicit connection in the \mm\ allows us to quantify how feedback affects the evolution of one of the most heavily irradiated clouds in Carina. 
In the following sections, we quantify the impact of external irradiation and compare it to the cold, molecular gas properties from \citet{reiter2023_MM}. 
The \mm\ is composed of multiple pillars within the larger complex. 
We refer to these pillars by the name of the jet they contain (i.e. the HH~901 pillar) as in \citet{reiter2023_MM}.

\subsection{Photoevaporation rate}\label{ss:photo_evap}

UV photons from Tr14 and indeed, the many other O- and B-type stars throughout the Carina star-forming complex irradiate and evaporate the surrounding molecular gas including the \mm.  
The large scale of the \mm\ complex means that the most heavily irradiated gas in the HH~901~pillar lies closer to Tr14, receiving $\sim 2$ times the radiation of more distant structures like the globule that hosts HH~1066. 
Overall, the UV irradiation of the \mm\ is roughly an order of magnitude higher than other well-studied pillars in Carina \citep[e.g.,][]{mcleod2016,klaassen2020}
More intense radiation appears to produce higher Mach numbers in the cold molecular gas of the \mm\ but not significantly higher turbulent compression \citep{menon2021,reiter2023_MM}.

The photoevaporative rocket effect has been proposed as one mechanism to drive the compression of pillars \citep[e.g.,][]{spitzer1954,lefloch1994,williams2001,arthur2011}. 
As fast electrons stream away from freshly ionized gas and into the H~{\sc ii} region, heavier positive ions are slowly pushed into the cloud by conservation of momentum. 
This compresses the gas, and possibly triggers its collapse. 
Multiple studies point to prominent jets like HH~901 and HH~902 that are found at the head of dust pillars as evidence of triggered star formation \citep[e.g.,][]{rag10,ohl12}. 
However, uncovered star formation is expected to produce many of the same observational signatures making claims of triggered star formation hard to verify \citep{dale2015}. 
The sub-arcsec view of the \mm\ provided by MUSE allows us to measure the photoevaporation rate along the irradiated interfaces. 
Comparing with ALMA observations of the cold, molecular gas allows us to connect the dots between the compression from the ionization front and the current physical state of the gas.

We measured the local photoevaporation rate at several ionized interfaces in the \mm, targeting the tips of pillars and bright ionized ridges that face Tr14 (shown as red circles in Figure~\ref{fig:sii_map}).
To compute the mass-loss rate at each location, we use the expression for a cylinder illuminated from one side from \citet{smith2004_finger} 
\begin{equation}
    \dot{M} \simeq 2 \pi r^2 m_{\mathrm{H}} n_{\mathrm{H}} v
\end{equation}
where 
$r$ is the radius of curvature,  
$m_{\mathrm{H}}$ is the mass of hydrogen, 
$n_{\mathrm{H}} = 1.4 n_e$ is the density of neutral gas, 
and 
$v$ is the velocity, assumed to be the sound speed in ionized gas ($v\approx 10$~km~s$^{-1}$). 
We determine the radius of curvature from the radius of the circle that best matches the curvature at the pillar head. 
We adopt a value for the sound speed, $v$, for an ideal isothermal gas: $v = \sqrt{\gamma k_B T / m}$ where 
$k_B$ is the Boltzmann constant, 
$m$ is the mean particle mass, 
and $\gamma$ is the adiabatic index \citep[5/3 assuming the H~{\sc ii} region is a monoatomic gas; see, e.g.,][]{spitzer1962}. 
For the range of $T_e$ found in this paper ($\sim 8000-10,500$~K), this corresponds to $v \approx 10-12$~\kms. 
The mass-loss rate at each interface is given in Table~\ref{t:props}. 
In the following sections, we consider how long the \mm\ will survive the ongoing evaporation and how much momentum this evaporation drives into the cold, molecular gas.

\subsection{Correlation with molecular gas properties}\label{ss:alma_comp}
\begin{figure}
    \includegraphics[width=\columnwidth]{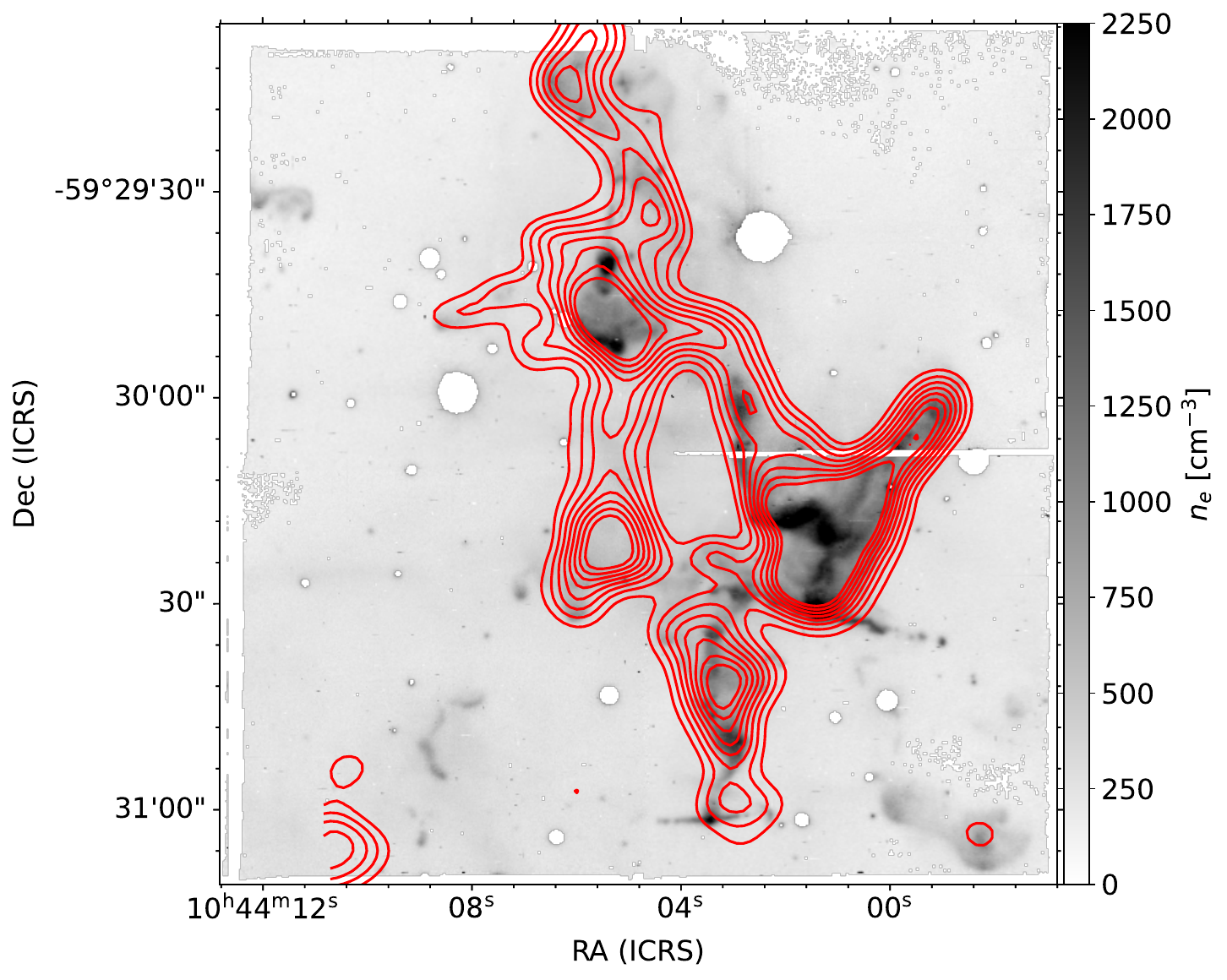}
    \includegraphics[width=\columnwidth]{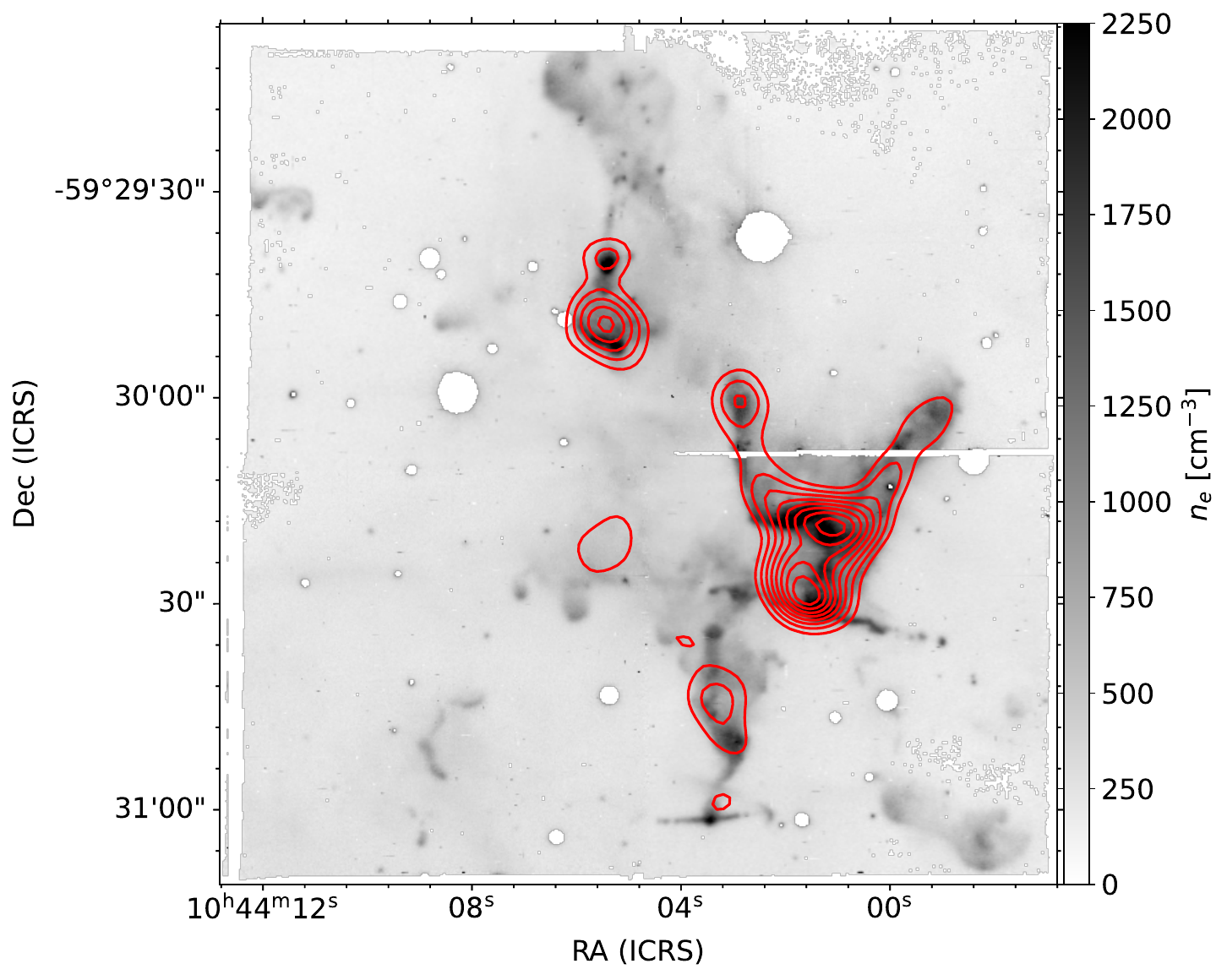}
    \includegraphics[width=\columnwidth]{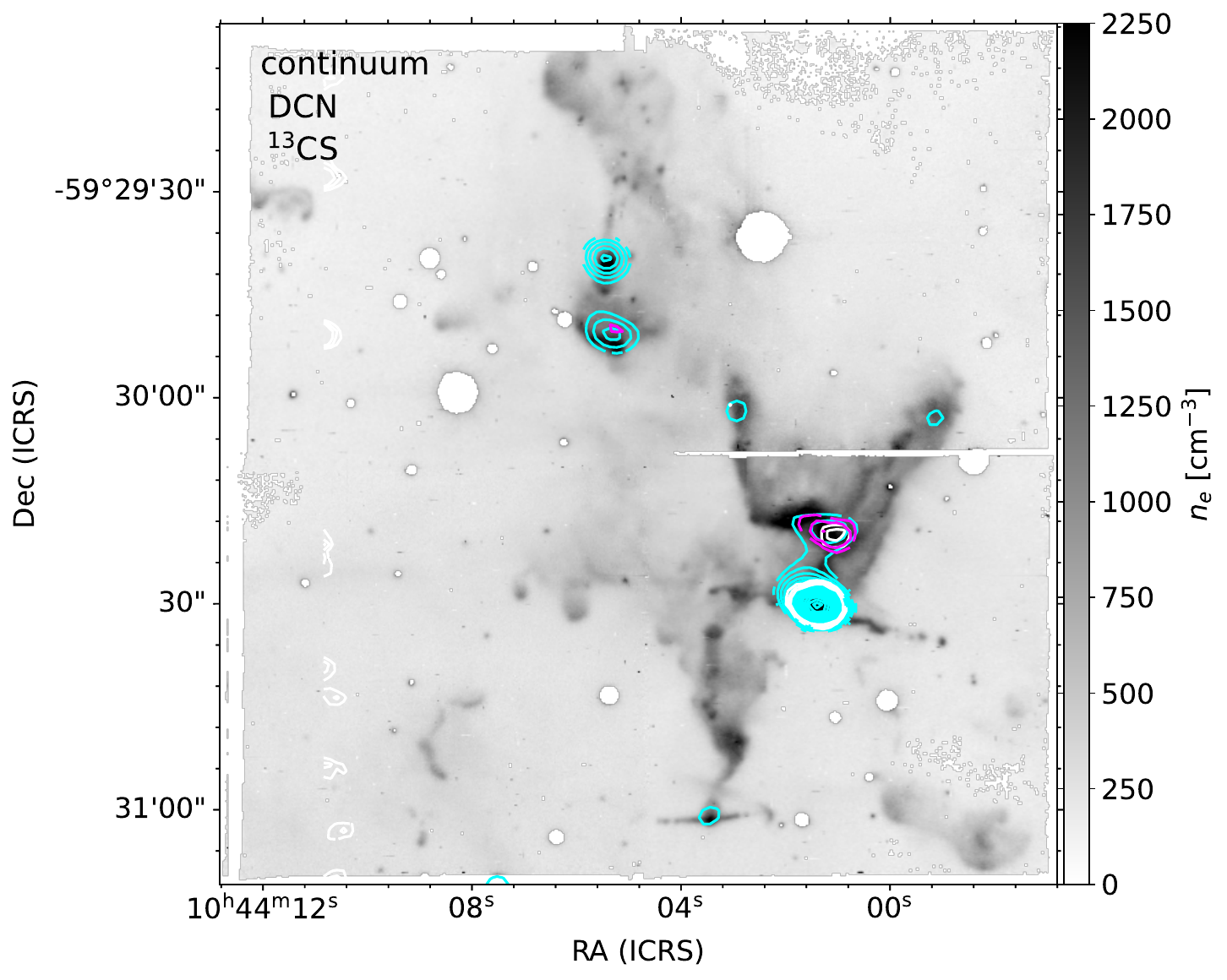}
    \caption{Electron density maps (grayscale) with contours of millimeter emission overlaid. 
    The top panel shows CO J=2-1 detected with $\geq5\sigma$. Contours show 20-100\% of the maximum emission in steps of 10\%. 
    The middle panel shows C$^{18}$O J=2-1 detected with $\geq5\sigma$ also with contours from 20-100\% in steps of 10\%, as for CO. 
    The bottom panel shows the 1~mm continuum (cyan; same contour levels as CO), DCN J=3-2 (white; contours are $3-5\sigma$ in steps of $1\sigma$), and $^{13}$CS J=5-4 (magenta; contours are $1-3\sigma$ in steps of $1\sigma$). 
    }
    \label{fig:density_comp}
\end{figure}

\subsubsection{CO and isotopologues}
Moderate resolution ($\sim 6$\arcsec) observations of the \mm\ revealed the large-scale structure of the cold molecular gas \citep{reiter2023_MM}. 
Contours showing the large-scale distribution of $^{12}$CO J=2-1 and C$^{18}$O J=2-1 from \citet{reiter2023_MM} are plotted on a map of $n_e$ in Figure~\ref{fig:density_comp}. 
These trace three distinct pillars in the \mm, each named for the famous jet it contains. 
A dendrogram analysis to identify the hierarchical  clustering in the gas identified these as three distinct structures in position-position-velocity space; these are shown as black contours in Figure~\ref{fig:ci_map}. 

In the \mm, 
\dco\ contours closely follow regions with high $n_e$. 
Both the HH~902 and the HH~1066 pillars have multiple \dco\ peaks, each corresponding to high density ridges or high curvature surfaces in the MUSE images. 
Broad, distinct ridges trace additional fragmentation within or in front of the pillar. 
The morphology of high $n_e$ surfaces in the HH~902~pillar suggests fragmentation into two wide pillars, each with a \dco\ clump at the head. 
The HH~1066 pillar is longer and narrower than the HH~902 pillar and appears to be fragmenting into multiple smaller globules. 
Two \dco\ clumps in the north of the HH~1066 pillar coincide with high $n_e$ ridges that face Tr14. 
A third, lower intensity \dco\ peak traces the southernmost tip of the HH~1066 pillar. 
Unlike the other \dco\ clumps, this southernmost clump does not lie immediately behind a high $n_e$ ionization front and there is no millimeter continuum detected at this position.

The correlation between \dco\ and $n_e$ does not extend to the HH~901 pillar. 
In the body of the HH~901 pillar, the brightest \dco\ emission is offset $\sim10$\arcsec\ north of the highest $n_e$ in the pillar. 
\dco\ contours lie slightly west of the high-density spine in the HH~901 pillar (see Figure~\ref{fig:density_comp}).

At the head of the HH~901 pillar, unresolved \dco\ more closely coincides with high electron densities near the HH~901~YSO. 
However, CO contours from the low-resolution observations fall slightly to the west of the pillar head (see Figure~\ref{fig:density_comp}). 
Higher resolution ($\sim 1$\arcsec) ALMA observations from \citet{cortes-rangel2020} reveal a band of diffuse emission in this area that extends west from the pillar and is unresolved in our beam. 
The velocity of this diffuse emission is redshifted $\sim 1.5$~\kms\ compared to the systemic velocity of the pillar \citep[$\sim -3.5$~\kms; see][]{reiter2023_MM}. 
Contamination from this diffuse emission may explain the offset seen at the pillar head in the low resolution ALMA map.

\subsubsection{Continuum and dense gas tracers}

Figure~\ref{fig:density_comp} shows 1.3~mm continuum contours (cyan) and the rarer isotopologues, $^{13}$CS J=5-4 (magenta) and DCN J=3-2 (white) on an $n_e$ map. 
All seven continuum peaks are found immediately behind bright ionization fronts. 
Five lie behind cloud edges that are perpendicular to the direction of radiation from Tr14. 
The two continuum sources seen at the northern edges of the HH~902 pillar lie inside ridges that are oriented nearly parallel to the radiation from Tr14, unlike the other five continuum sources. 
The pillar walls at this location have the lowest curvature of the ionized interfaces in the \mm. 
Estimated masses of both clumps are $\lesssim 0.5$~\Msun, smaller than the other clumps in the HH~902 pillar ($\sim 1.2$~\Msun\ and $\sim 5.3$~\Msun), more similar to the low mass of HH~901~mm.

$^{13}$CS J=5-4 traces high density gas and is primarily seen in the clump behind the HH~902~c ionized ridge. 
This source and the clump immediately south of it are both detected in DCN J=3-2. 
Deuterated molecules like DCN are quickly destroyed by CO in the gas phase so tend only to be found in cold gas \citep[e.g.,][]{dalgarno1984,caselli2012,ceccarelli2014}. 
High densities allow CO to self-shield and freeze out onto grains. 
Both clumps with a DCN detection lie behind some of the highest electron densities in the pillars ($\gtrsim 2000$~cm$^{-3}$, see Table~\ref{t:props}). 
Other regions with similarly high densities do not have an associated DCN detection (HH~1066~a) although this may be due to beam dilution. 
Indeed, higher resolution ($\sim1$\arcsec) observations from \citet{cortes-rangel2020,cortes-rangel2023} reveal N$_2$D$^+$ J=3-2 in all of the targeted YSOs in the \mm\ (the HH~901, HH~902, and HH~1066 YSOs).

\subsection{The remaining lifetime of the \mm\ }\label{ss:lifetime}

\begin{table}
	\centering
	\caption{Molecular gas mass of the \mm\ and the smaller pillars it contains. Data from \citet{reiter2023_MM}. Pillar areas shown in Figure~\ref{fig:ci_map}. }
	\label{t:mol_masses}
	\begin{tabular}{lccl} 
		\hline
		Pillar & mass & area &  comment \\
        name & [\Msun] & [pc$^{-2}$] & \\ 
		\hline
		\mm\ & $33.7 \pm 5.7$ & $0.32\pm 0.03$ & entire complex \\
		HH~901 & $3.31 \pm 2.64$ & $0.07 \pm 0.007$ & does not include tip \\
		HH~902 & $17.2 \pm 13.2$ & $0.18 \pm 0.02$ &  \\
        HH~1066 & $8.13 \pm 7.51$ & $0.20 \pm 0.02$ &  \\
		\hline
	\end{tabular}
\end{table}

With the photoevaporation rates computed in Section~\ref{ss:photo_evap} and the mass ($M = 33.7 \pm 5.7$~\Msun; see Table~\ref{t:mol_masses}) from \citet{reiter2023_MM}, we can estimate the remaining \mm\ lifetime. 
For the average photoevaporation rate, $\sim 4 \times 10^{-6}$~\Msun~yr$^{-1}$, the entire \mm\ will be completely evaporated in $8 \pm 2$~Myr. 
However, photoevaporation rates vary by two orders of magnitude within the \mm, with lifetimes as short as $\sim 1$~Myr in some sub-regions (using masses from Table~\ref{t:mol_masses} and photoevaporation rates from Table~\ref{t:props}). 
\citet{cortes-rangel2020} find even shorter remaining lifetimes, $\sim 10^4 - 10^5$~yr, for the molecular cocoons surrounding HH~901 and HH~902.

If different parts of the complex are indeed evaporated more quickly, then the forming stars will be exposed at different times.  
Unsurprisingly, the lowest photoevaporation rate is in the HH~1066 globule, which has the largest separation from Tr14 ($\sim 6 \times 10^{-7}$~\Msun~yr$^{-1}$). 
However, the most highly irradiated gas at the head of the HH~901 pillar has a photoevaporation rate that is only a factor of $\sim 2$ higher, despite the close proximity to Tr14. 
The highest photoevaporation rates are found in the largest ionization fronts (e.g., in the HH~902 pillar, see Table~\ref{t:props}).

Photoevaporation rates are lower in the smaller structures in the \mm. 
Both the HH~901 and HH~1066 YSOs reside in pillar heads with a small radius of curvature, 0.01~pc (see Table~\ref{t:props}). 
However, the HH~1066~YSO is detectable in the near-IR \citep{reiter2014,reiter2016} while the HH~901~YSO has only been detected at 1~mm with ALMA \citep{cortes-rangel2020}. 
The ionization front at the head of the HH~901 pillar is smaller than that ahead of the HH~1066~YSO 
(HH~1066~a, see Figure~\ref{fig:ifronts}),
indicating that ionizing photons do not penetrate as deeply into the cloud. 
This suggests that higher UV irradiation may have compressed the HH~901 pillar, driving the gas to higher densities which may enhance the shielding of the embedded YSO. 
For a remaining envelope mass of 0.52~\Msun\ \citep[estimated from the 1.3~mm continuum; see][]{reiter2023_MM}, the HH~901~YSO will remain shielded for $\gtrsim2.5$~Myr at the current photoevaporation rate.

\subsection{The erosion of the \mm\ }

Several authors have taken the presence of prominent jets emerging from the heads of dust pillars as evidence of triggered star formation \citep[e.g.,][]{billot2010,rag10,ohl12,ohlendorf2013,panwar2019}. 
However, pre-existing overdensities may halt erosion locally, leading to the formation of pillars, including the observed pillars-within-pillars structure of the \mm. 
Unambiguous evidence of triggered star formation is difficult to identify in observations and simulations suggest that there are few physical differences between triggered and non-triggered stars \citep{dale2015}. 

The morphology in the MUSE images suggests that the HH~901 and HH~1066 YSOs reside in globules that are in the process of detaching from their parent pillars. 
North of the HH~901~YSO, the pillar width tapers to $\sim 1$\arcsec, half the width at the pillar head. 
Electron densities are $\sim 1/3 - 1/2$ lower in the narrower `neck' of the pillar compared to regions immediately north and south of it. 
Another high $n_e$ region $\sim 12$\arcsec\ north of the HH~901~YSO traces what will be the pillar head once the globule detaches.

The HH~901 globule resembles another irradiated globule also in the center of the Carina star-forming complex \citep{reiter2020_combined}. 
The so-called tadpole globule is bombarded with UV photons from nearby Tr16. 
The globule hosts a deeply embedded YSO that has only been detected at 1.3~mm with ALMA \citep{reiter2020_alma,cortes-rangel2023} that drives the HH~900 jet+outflow. 
Both the tadpole and the HH~901 globules have large molecular reservoirs such that the YSO and its planet-forming disk will be shielded for $>$1~Myr, allowing ample time for planet formation to proceed with limited impact from the nearby high-mass stars.

Free floating globules like the tadpole and, in the future, the HH~901 globule, have been proposed as precursors to proplyds \citep[e.g.,][]{schneider2016}. 
In this case, the remaining envelope shields the disk as the region dynamically evolves so that a star may arrive in the high UV fields in the center of the cluster with a robust disk. 
However, this is unlikely to be the fate of the HH~901 globule.
External UV irradiation from Tr14 irradiates the pillar on one side, driving a shock into the gas via the rocket effect that will compress and accelerate the globule \citep[e.g.,][]{mellema1998}. 
This will produce a net velocity \emph{away} from the ionizing sources, as seen in the tadpole \citep{reiter2020_combined}. 
The velocity of the tadpole away from the ionizing sources ($\sim 10$~\kms) is larger than can be explained by turbulent fragmentation, pointing toward possible triggered star formation.

Unlike HH~900, the existing velocity measurements of the HH~901 pillar do not require an external accelerant. 
The radial velocity of the globule is comparable to the pillar velocity \citep{cortes-rangel2020}. 
Motion primarily in the plane of the sky may explain the lack of radial velocity difference. 
However, if this is the case, we expect a visible bend in the jet toward the ionizing sources. 
Instead, the HH~901 jet bends away from Tr14, perhaps because the jet itself has been accelerated by the rocket effect \citep{reiter2013}. 
For HH~901, we cannot rule out the possibility that this is an example of uncovered star formation, where 
the nascent YSO inherited the velocity of the cloud before external UV irradiation began accelerating the gas.

The (detaching) HH~1066 globule provides another example. 
The larger HH~1066 pillar appears to be fragmenting into multiple globules. 
Two \dco\ clumps in the north of the HH~1066 pillar coincide with features that look like globules in the MUSE images. 
Both have curved surfaces with high electron densities along the surface facing Tr14. 
A tear-drop-shaped tail extends from the northernmost globule, leading this to be identified as a candidate proplyd in earlier ground-based images \citep{smith2003}. 
However, the large gas mass and the presence of the HH~1066 jet and YSO demonstrate that this is more akin to an evaporating gaseous globule \citep[e.g.,][]{sahai2012}.

Unlike the HH~900 and HH~901 YSOs, the HH~1066~YSO is detected at $\lambda < 1$~\micron\ (including the reddest wavelengths of the MUSE coverage). 
Taken at face value, this suggests that the HH~1066~YSO is more evolved. 
However, it may instead be embedded in a less dense globule. 
Unlike HH~900 and HH~901, the HH~1066 jet emerges from the globule immediately \citep{reiter2013}, with shock knots seen just above and below the marginally resolved circumstellar disk \citep{mesa-delgado2016}. 
Radial velocities in the HH~1066 jet (see Section~\ref{ss:jets}) suggest a larger tilt away from the plane of the sky so it is also possible that the more evolved appearance of the HH~1066~YSO reflects a viewing angle down the outflow cavity. 
Unlike HH~901, the HH~1066 globule is slightly blueshifted ($\sim 3-5$~\kms) compared to nearby pillar gas \citep{cortes-rangel2023} and the jet bends slightly toward Tr14 \citep{reiter2017}. 

Taken together, the HH~900, 901, and 1066 globules provide mixed evidence for triggered star formation. 
However, the \mm\ points to another important way that the presence of feedback may affect stars unequally. 
High density gas at the tips of the pillars will shield the embedded YSOs from the harsh UV environment for $>$1~Myr, providing ample time for planet formation to proceed unimpeded. 
When and how intensely disks are irradiated determines the mass, orbital architecture, and chemistry of planets as these three factors are all strongly affected by external UV radiation \citep[e.g.,][]{berne2023,qiao2023,wilhelm2023}. 
\citet{qiao2023} showed that a shielding time of $\sim 1.5$~Myr is required to nullify the impact of external photoevaporation on pebble accretion in planet-forming disks. 
Shorter shielding times lead to a reduction in the pebble reservoir in the outer disk, reducing the mass and orbital radii of the planets formed  \citep{qiao2023,wilhelm2023}. 
On a population level, the fraction of stars shielded as a function of time is poorly constrained. 
\citet{qiao2022} estimated relatively brief ($< 0.5$~Myr) shielding times for only a small fraction of disks in a Carina-like region. 
The \mm\ provides one example of how the shielding time can vary in a small region of highly structured gas. 
A more complete observational test requires quantifying the gas clearing time as a function of the incident radiation and over a wider variety of structures.

\subsection{Photoevaporating outflows}\label{ss:outflows}

The prominent outflows seen from the \mm\ all have relatively young dynamical ages of a few hundred (HH~1066) or a few thousand (HH~901 and HH~902) years \citep{reiter2017}. 
Taken at face value, this suggests that all three jet-driving YSOs are young and only recently in an active outflow-driving stage. 
However, once in the H~{\sc ii} region, each outflow is heavily irradiated with UV photons from Tr14. 
This not only illuminates the jet body but evaporates it. 
Jet knots from older ejecta may no longer be visible / identifiable because they have been completely evaporated. 
This provides one possible explanation for why the outflow with the youngest dynamical age (HH~1066) appears most evolved because it is detected at wavelengths shorter than 1~mm.

External illumination and evaporation also change the appearance of the outflows. 
Inside the dust pillars, HH~901 and HH~902 are seen as cold molecular CO outflows that emerge from deeply embedded YSOs \citep{cortes-rangel2020}; molecular emission from the HH~1066 outflow could not be disentangled from the ambient molecular gas \citep{cortes-rangel2023}.  
No CO emission survives outside the protection of the dust pillar. 
Instead, ionized gas tracers like H$\alpha$ trace the length of the outflow \citep{smith2010}. 
[Fe~{\sc ii}] emission traces the densest portions of the jets in the H~{\sc ii} region \citet{reiter2013}.

In HH~901, the emission structure traces the dissociation of the outflow as it enters the H~{\sc ii} region. 
Some extended molecular emission along the outflow axis was already identified in near-IR H$_2$ images from \citet{hartigan2015}. 
Extended [C~{\sc i}] $\lambda 8727$~\AA\ emission seen with MUSE 
traces approximately the same morphology as the H$_2$.
We marginally resolve a velocity shift between the [C~{\sc i}] in the eastern and western jet limbs (see Figures~\ref{fig:hh901_ci} and \ref{fig:hh901_jet_vels}). 
[Fe~{\sc ii}] emission from the collimated jet is offset from the globule edge, becoming bright only where the [C~{\sc i}] emission ends (see Figure~\ref{fig:hh901_jet_vels}). 
This same structure is seen in HH~900, another externally irradiated outflow in Carina \citep{reiter2015_hh900,reiter2019_tadpole,reiter2020_alma,cortes-rangel2020}.
High densities in both outflows allow molecules to survive briefly in the H~{\sc ii} region. 
Once they are dissociated, optical depths drop enough to allow UV photons with $\geq$7~eV to ionize Fe in the high-density jet core. 
Both HH~902 and HH~1066 are more difficult to distinguish from the bright, irradiated pillar gas, so it is unclear if they also have a similarly resolved dissociation structure. 

The most unambiguous examples of dissociating molecular outflows, HH~900 and HH~901, are among the most heavily irradiated outflows.
Both emerge from compact, high-density globules. 
It is unclear why the outflows with a large enough column of molecules to be seen even briefly in the H~{\sc ii} region come from the most compact (and most heavily irradiated) globules. 
One possibility is that the compression of the globules is ongoing and this forces fresh material into the outflow path. 
In a quiescent environment, this outflow path would remain cleared and probably grown with time. 
Higher spatial resolution observations are required to determine if, for example, the HH~1066 outflow also has spatially resolved [C~{\sc i}] or H$_2$ emission tracing the dissociation of the molecular outflow.

\section{Conclusions}

In this paper, we present MUSE observations of the \mm, a heavily irradiated set of dust pillars in the heart of the Carina Nebula. 
Copious UV photons from Tr14 illuminate the pillars with an order of magnitude more intense radiation than sampled in other studies of pillars in Carina.  
We use the spatial and spectral coverage of MUSE to measure the density, temperature, excitation, and ionization of the hot ionized gas as a function of position. 
High electron densities trace pillar tips and ionization fronts where the \mm\ appears to be fragmenting into smaller pillars and globules. 
Intensity tracings across ionized interfaces in the \mm\ reveal smaller ionization fronts ahead of dense clumps and embedded YSOs. 

The MUSE data provide an excellent view of the famous jets, HH~901, 902, and 1066, that emerge from the pillar tips. 
HH~901 and 902 stand out in maps of the electron density and ionization, revealing high-density, low-ionization gas throughout the length of the jets; HH~1066 is more difficult to distinguish from the pillar ionization front. 
We extract position-velocity diagrams for all three jets in lines tracing different excitation from highly ionized (H$\alpha$) to low ionization ([S~{\sc ii}] and [Fe~{\sc ii}]) and neutral ([O~{\sc i}], [C~{\sc i}]) gas. 
Radial velocities are only marginally resolved for HH~901 and 902 which lie close to the plane of the sky while HH~1066 shows a clear red-and blueshift in [Fe~{\sc ii}] and [O~{\sc i}]. 
Only HH~901 shows extended [C~{\sc i}] emission, likely tracing the dissociation of the molecular outflow in the H~{\sc ii} region. 
We also identify a new candidate jet to the north of HH~902 seen only in [Fe~{\sc ii}]. Proper motion measurements in the future may confirm the jet-like nature of this source.

One of the goals of this work is to compare the ionized gas properties seen with MUSE to the structure and kinematics of the cold molecular gas as seen with modest (6\arcsec) resolution by ALMA.  
Cold, dense gas traced by \dco\ peaks behind ionization fronts and is closely correlated with high electron densities in the MUSE maps. 
One notable exception is the HH~901 pillar where the \dco\ peak is offset $\sim10$\arcsec\ from the electron density peak. 
We use the physical properties of the ionized gas derived from the MUSE data to calculate the photoevaporation rate in the \mm. 
Combining this photoevaporation rate with the molecular gas mass measured with ALMA, we calculate the remaining lifetime for the \mm\ as a whole of $8 \pm 2$~Myr.
However, when we consider the individual pillars and globules within the \mm, the estimated lifetimes vary from $\sim 1$~Myr to $\sim 13$~Myr. 
The longest lifetimes are for the smallest structures, independent of their relative distance from the ionizing sources. 
For example, the tip of the HH~901 pillar is closest to Tr14 but because of its small size, the remaining globule lifetime is estimated to be $\gtrsim$2.5~Myr.

Variations in the photoevaporation rate over the \mm\ suggests that the embedded YSOs will be revealed at different times. 
High densities and small sizes of pillar heads suggest that they may have been compressed by external UV irradiation from Tr14. 
This is consistent with \citet{reiter2023_MM} who found evidence for compressive turbulence in the \mm. 
Gas velocities from ALMA and jet bending do not clearly point to an acceleration of the gas and stars forming within it, as expected for triggered star formation. 
In-hand data are consistent with star formation being revealed as the cloud is evaporated. 
Density enhancements that are surrounded by ionization fronts deeper in the pillars, away from current pillar heads, suggests that erosion and fragmentation are ongoing.

However, the compression of pillars and globules may have enhanced their remaining lifetime. 
Star-forming cocoons may shield the protostar and its disk from the harsh environment for a significant fraction of the planet-formation timescale. 
Recent theoretical studies point to the shielding time as a key variable in planet formation. 
External photoevaporation destroys disks from the outside in, starving forming planetesimals from pebbles from the outer disk, limiting their final masses and orbital architectures. 
Measuring the photoevaporation rate and shielding time in high-mass star-forming regions like Carina is key to constrain the role of shielding in planet formation.

\section*{Acknowledgements}

This paper is based on data obtained with ESO telescopes at the Paranal Observatory under programme ID 097.C-0137 (PI: A.~F.\ McLeod). 
This paper makes use of the
following ALMA data: ADS/JAO.ALMA\#2018.1.01001.S.
ALMA is a partnership of ESO (representing its member
states), NSF (USA) and NINS (Japan), together with
NRC (Canada) and NSC and ASIAA (Taiwan) and KASI
(Republic of Korea), in cooperation with the Republic of
Chile. The Joint ALMA Observatory is operated by ESO,
AUI/NRAO and NAOJ. 
The National Radio Astronomy Observatory is a facility of the National Science Foundation operated under cooperative agreement by Associated Universities, Inc.
This research has made use of the SIMBAD database,
operated at CDS, Strasbourg, France \citep{wenger2000}. 
DI acknowledges support from the European Research Council (ERC) via the ERC Synergy Grant ECOGAL (grant 855130) and from collaborations and/or information exchange within NASA’s Nexus for Exoplanet System Science (NExSS) research coordination network sponsored by NASA’s Science Mission Directorate under Agreement No. 80NSSC21K0593 for the program “Alien Earths”.

\section*{Data Availability}

The MUSE data presented in this paper are publicly available from the ESO archive\footnote{\href{https://archive.eso.org/cms.html}{https://archive.eso.org/cms.html}}. 
We are happy to share reduced data products on request. 
The ALMA data used in this study are publicly available from the ALMA archive \footnote{\href{https://almascience.nrao.edu/aq/?result_view=observations}{https://almascience.nrao.edu/aq/?result\_view=observations}} under the program ID number ADS/JAO.ALMA\#2018.1.01001.S.



\bibliographystyle{mnras}
\bibliography{bibliography} 




\appendix

\section{$A_V$ maps}\label{A:AV}

Bright stars and bright nebular lines like H$\alpha$ are saturated in the long exposures but not in the short exposures. 
We therefore use a combination of the two maps to measure the extinction in the \mm\ region.

\begin{figure}
    \includegraphics[width=\columnwidth]{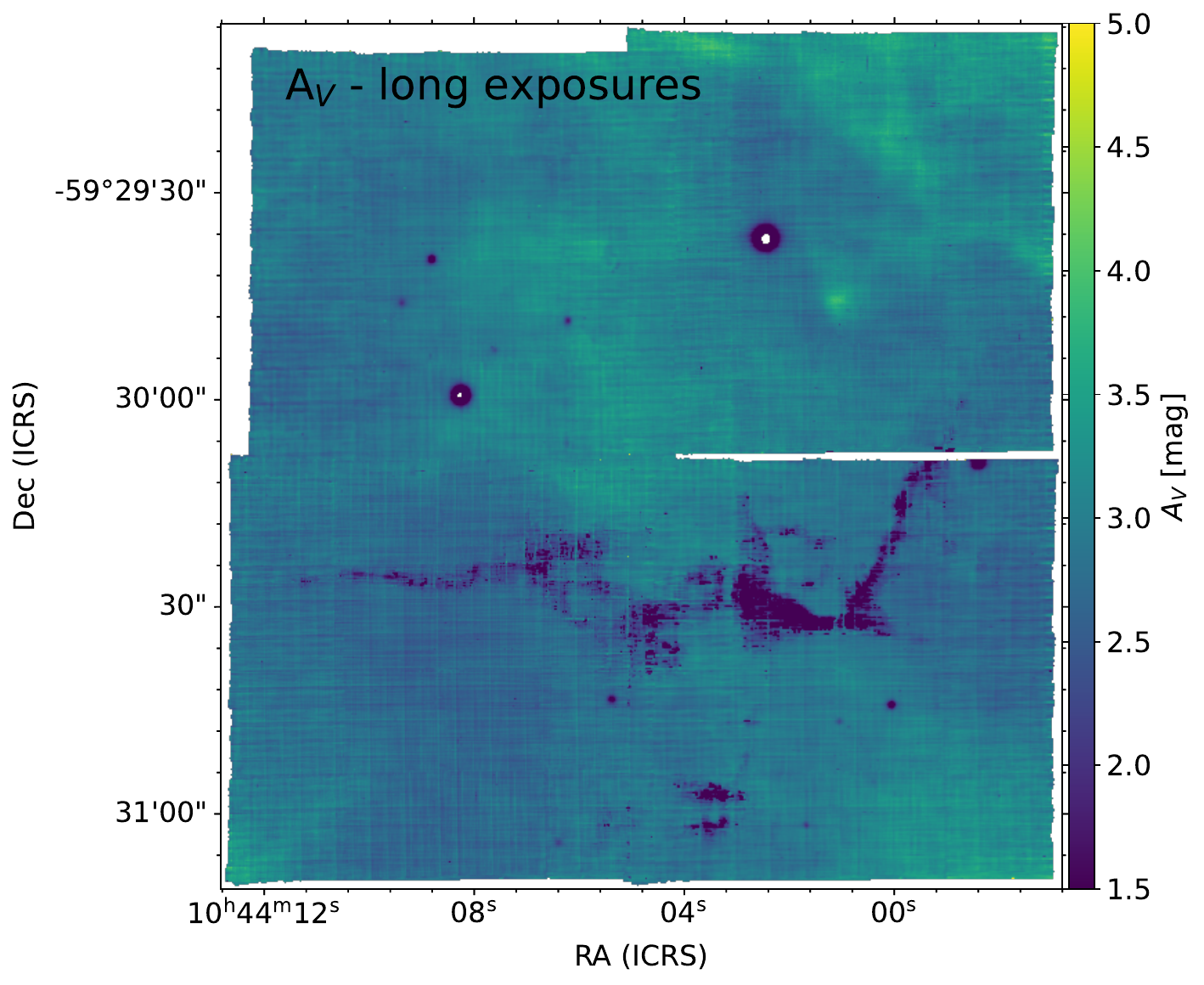}
    \includegraphics[width=\columnwidth]{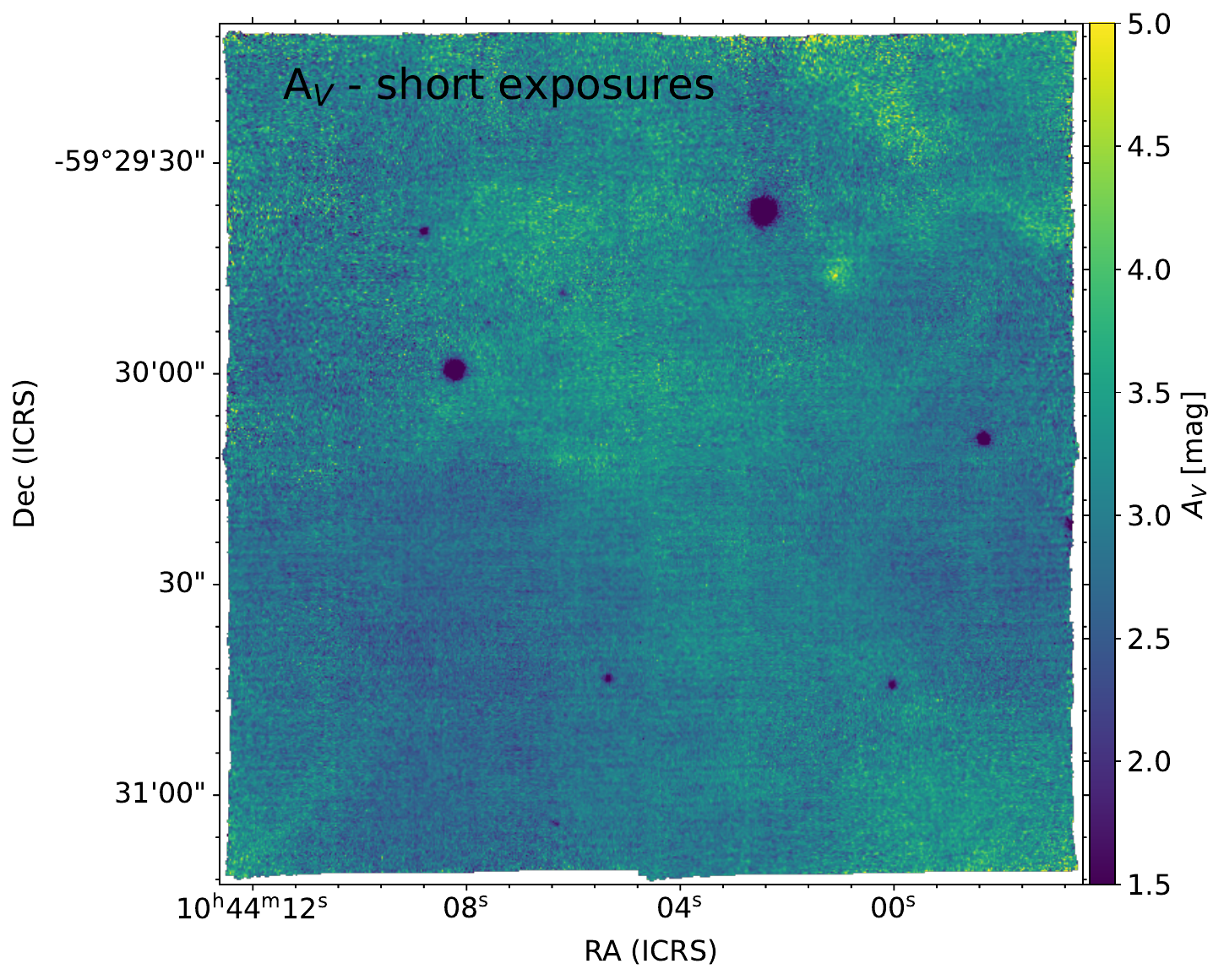}
    \includegraphics[width=\columnwidth]{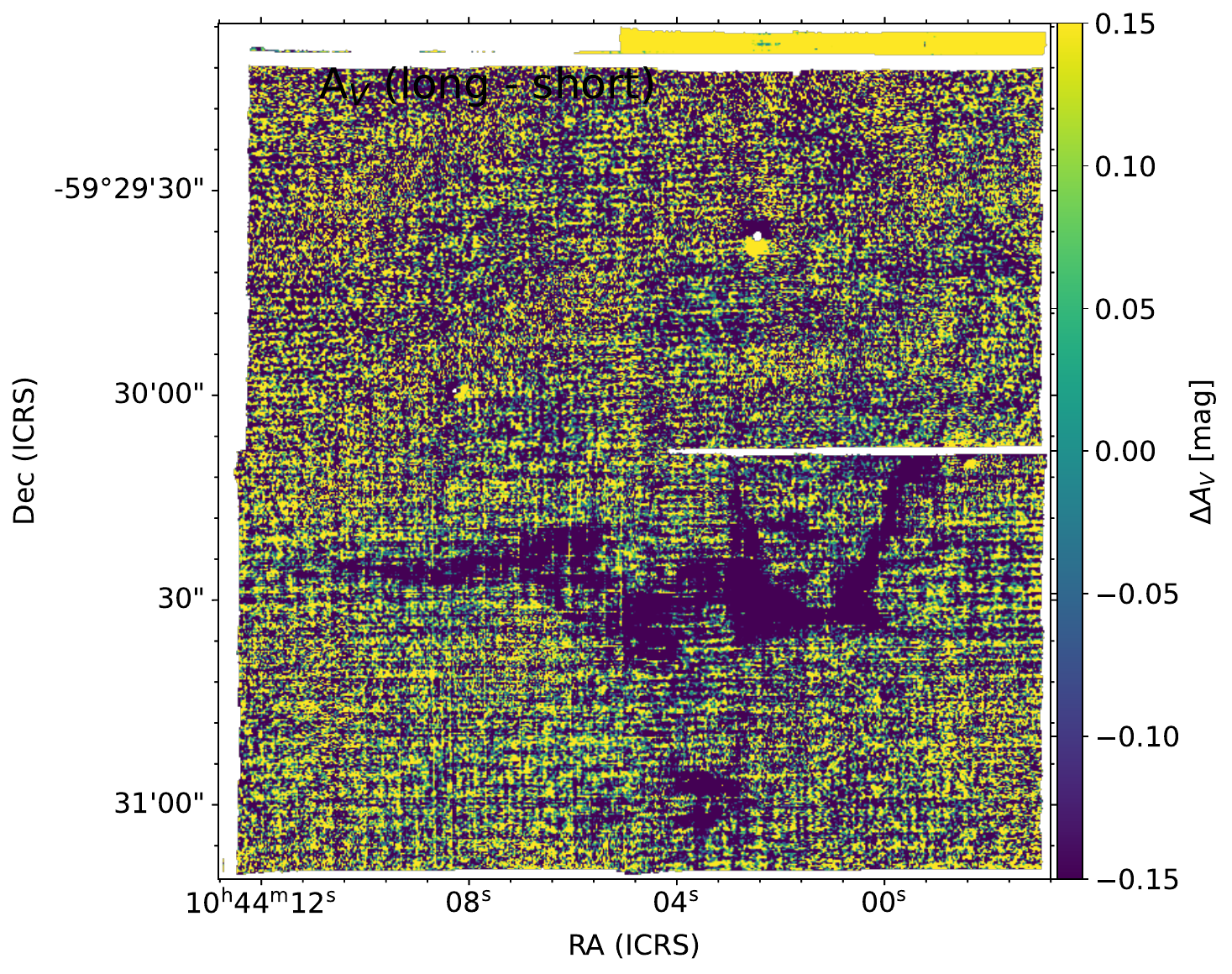}
    \caption{Extinction maps computed from the long (\textbf{top}) and short (\textbf{middle}) exposures. The \textbf{bottom} panel shows the difference of the two maps showing that they differ by more than 0.1~mag only where the long exposures are saturated on bright stars or bright ionization fronts.  
    }
    \label{fig:AV_maps}
\end{figure}

\section{Jet position-velocity diagrams}\label{A:jet_PVs}

\begin{figure}
	\includegraphics[width=\columnwidth]{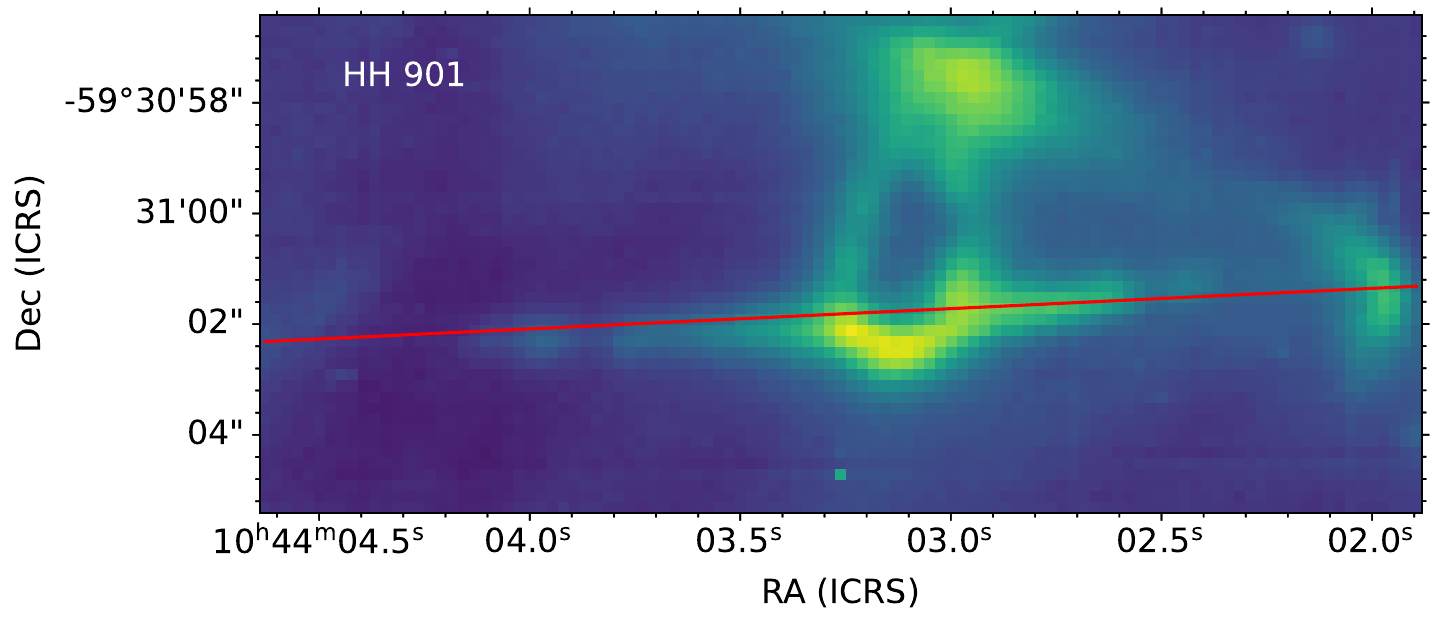}
    \includegraphics[width=\columnwidth]{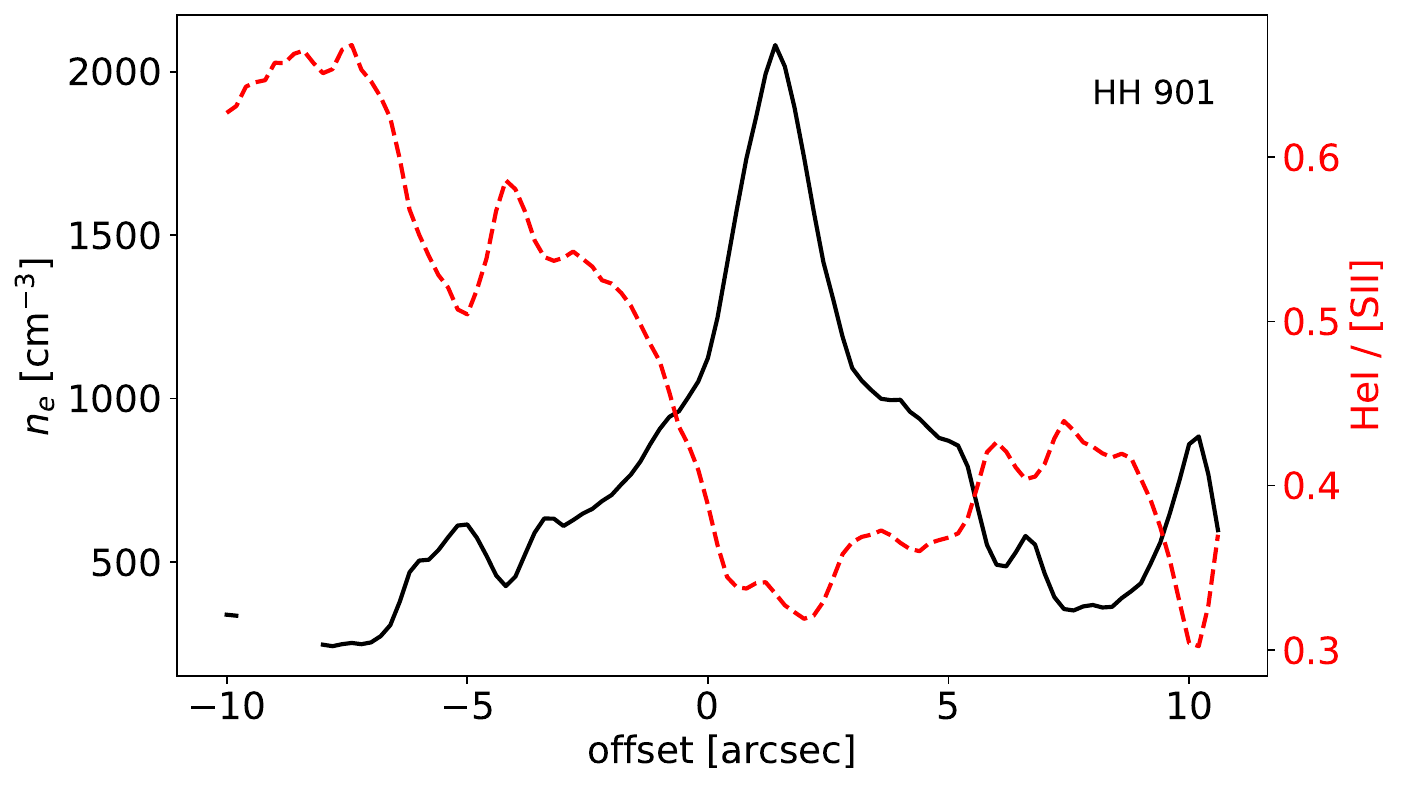}
    \includegraphics[width=\columnwidth]{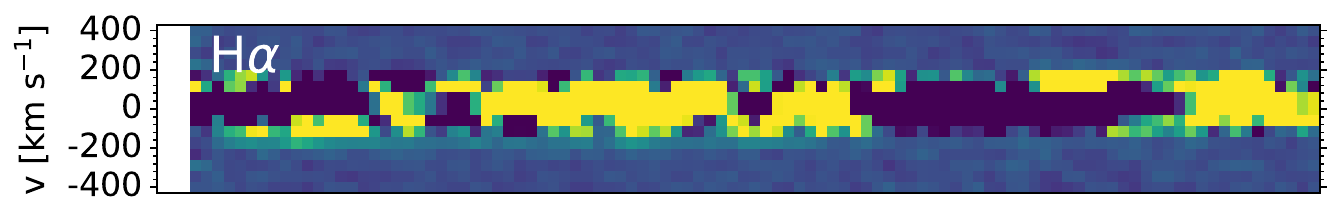}
    \includegraphics[width=\columnwidth]{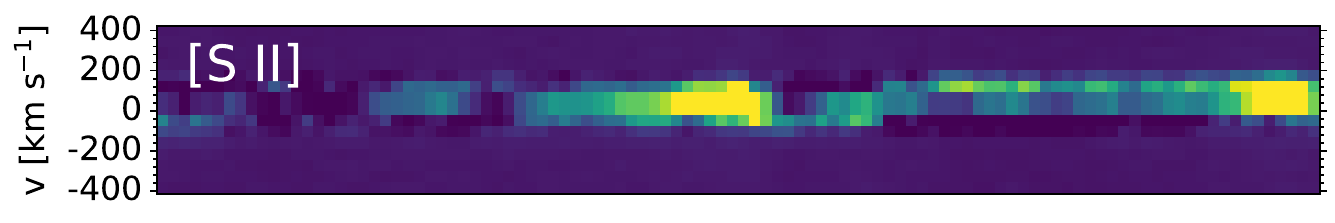}
    \includegraphics[width=\columnwidth]{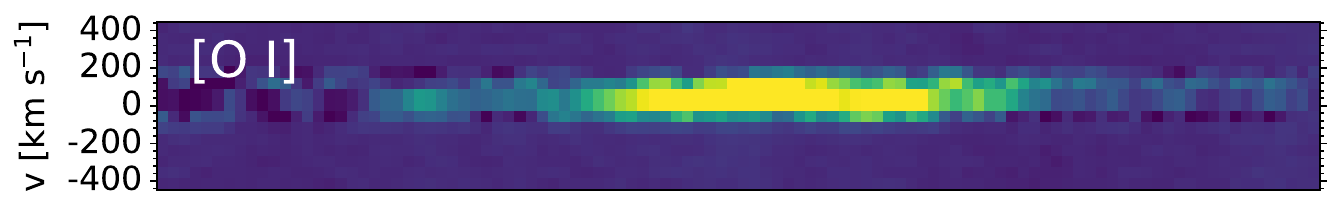}
    \includegraphics[width=\columnwidth]{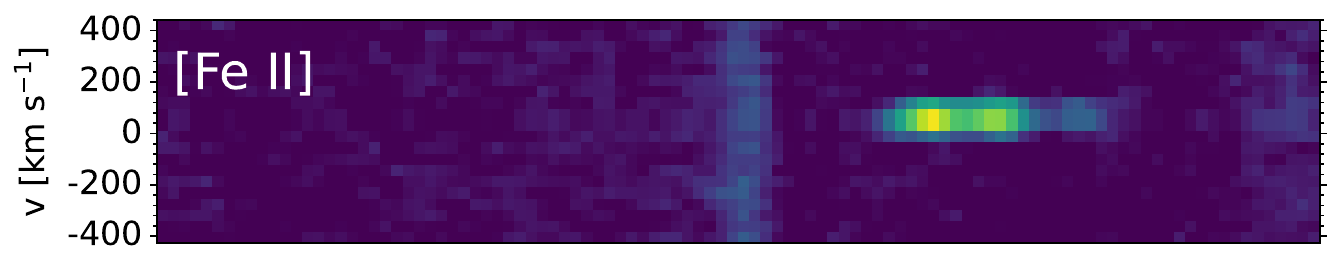}
    \includegraphics[width=\columnwidth]{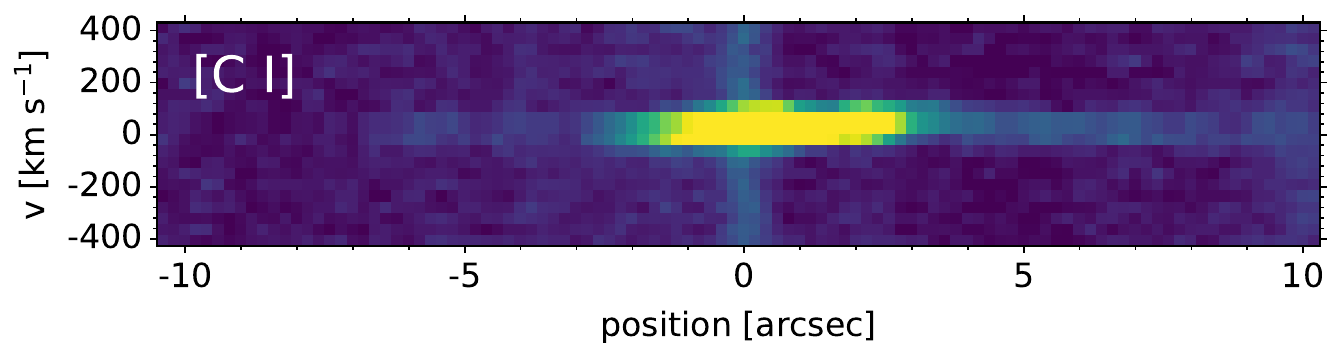}
    \caption{ Position-velocity diagrams of HH~901. 
    Top panel shows the location of the P-V extraction on an [S~{\sc ii}] image. Second panel shows the density profile along the jet (black) and the degree of ionization along the jet from the ratio He~{\sc i} $\lambda 6678$~\AA\ / [S~{\sc ii}] $\lambda 6717, 6731$~\AA\ (red, dashed line). P-V diagrams below that are: H$\alpha$; [S~{\sc ii}] $\lambda 6717$~\AA; [O~{\sc i}] $\lambda 6300$~\AA; [Fe~{\sc ii}] $\lambda 8616$~\AA; and [C~{\sc i}] $\lambda 8727$~\AA.}
    \label{fig:hh901_jet_vels}
\end{figure}

\begin{figure*}
	\includegraphics[width=\columnwidth]{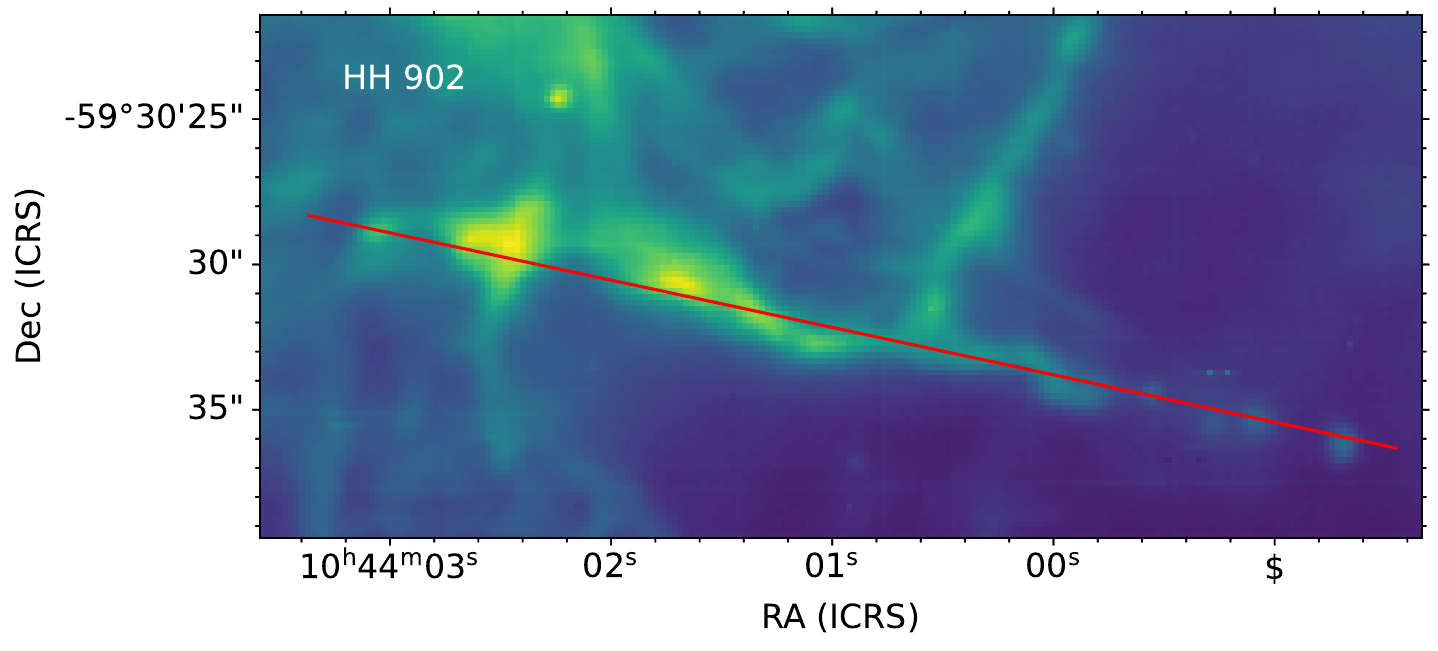}
    \includegraphics[width=\columnwidth]{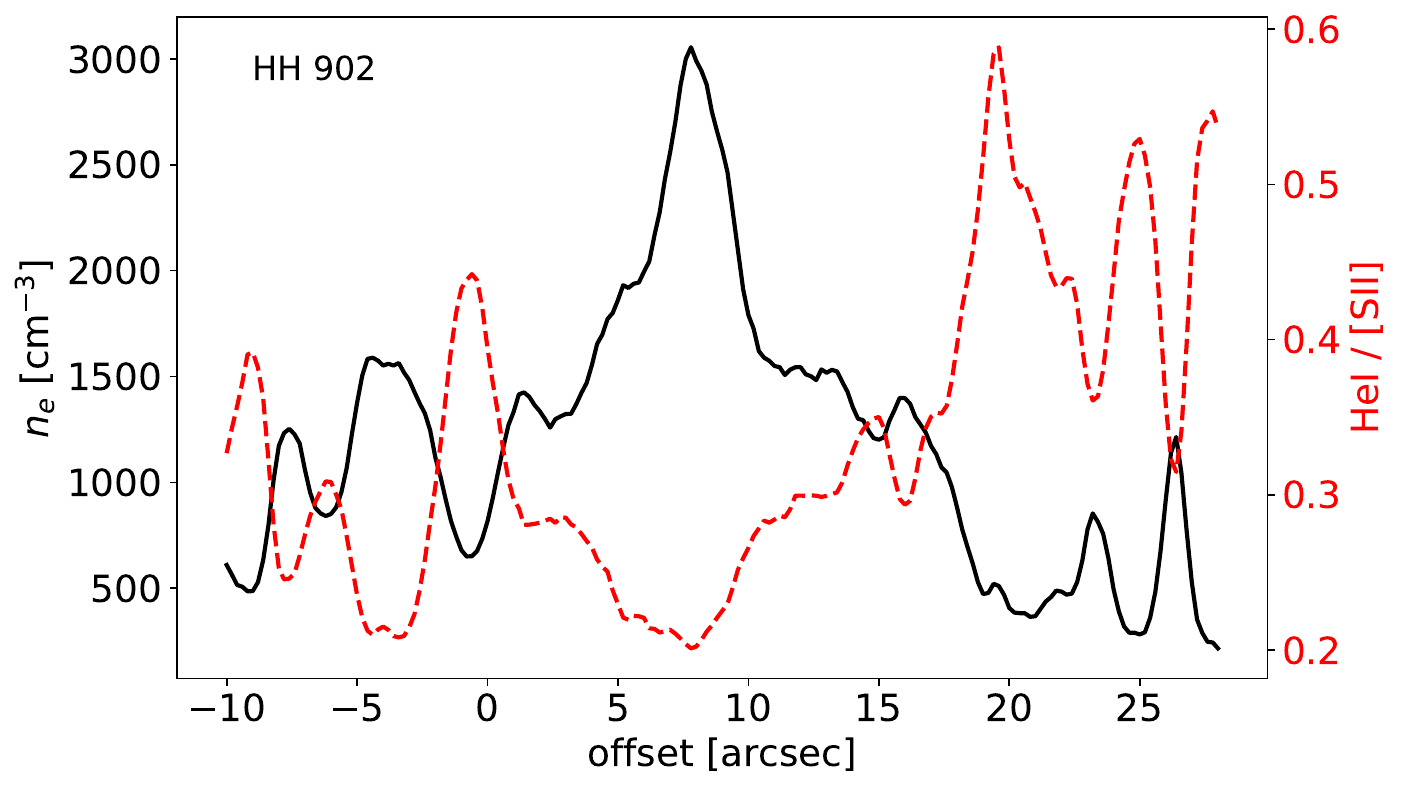}
    \includegraphics[width=\textwidth]{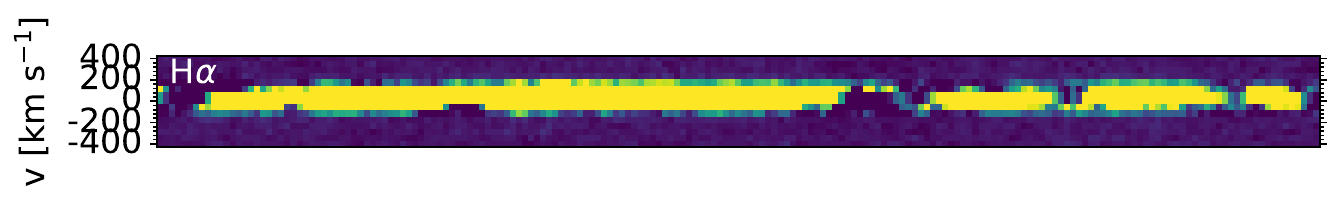}
    \includegraphics[width=\textwidth]{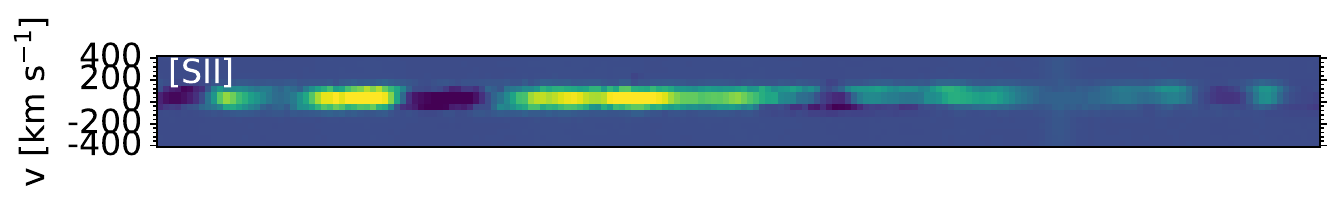}
    \includegraphics[width=\textwidth]{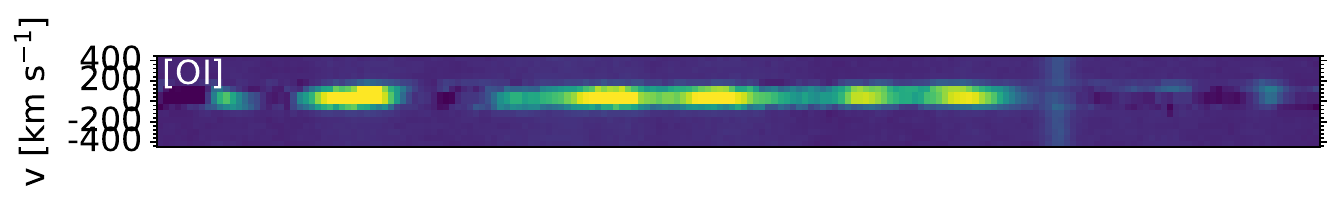}
    \includegraphics[width=\textwidth]{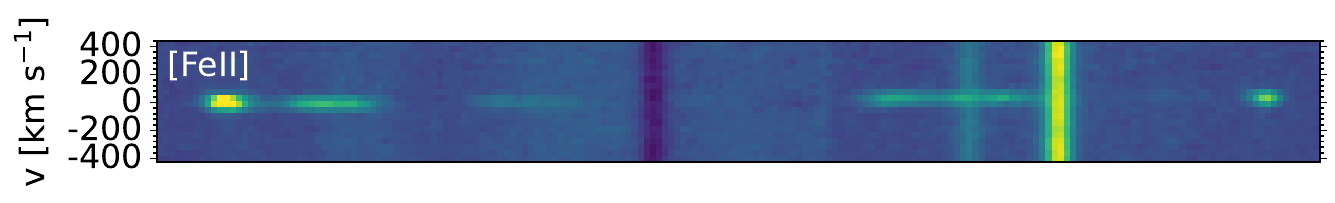}
    \includegraphics[width=\textwidth]{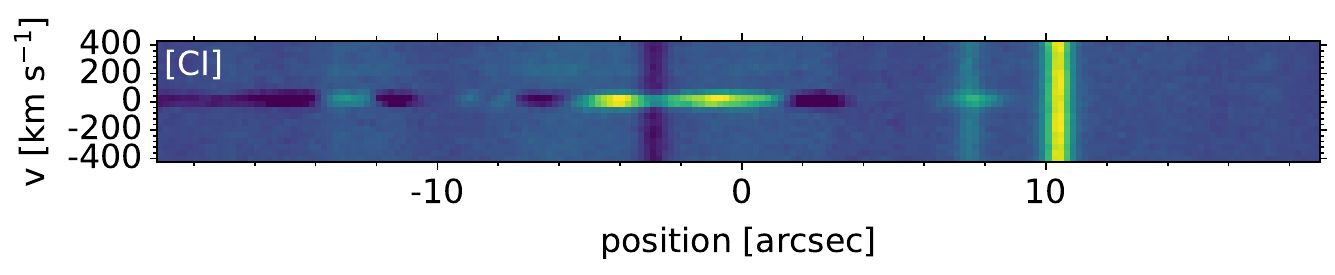}
    \caption{ Same as Figure~\ref{fig:hh901_jet_vels} but for HH~902. Two continuum sources can be seen at $\sim 7.5$\arcsec\ and $\sim 10.5$\arcsec.}
    \label{fig:hh902_jet_vels}
\end{figure*}

\begin{figure*}
	\includegraphics[width=\columnwidth]{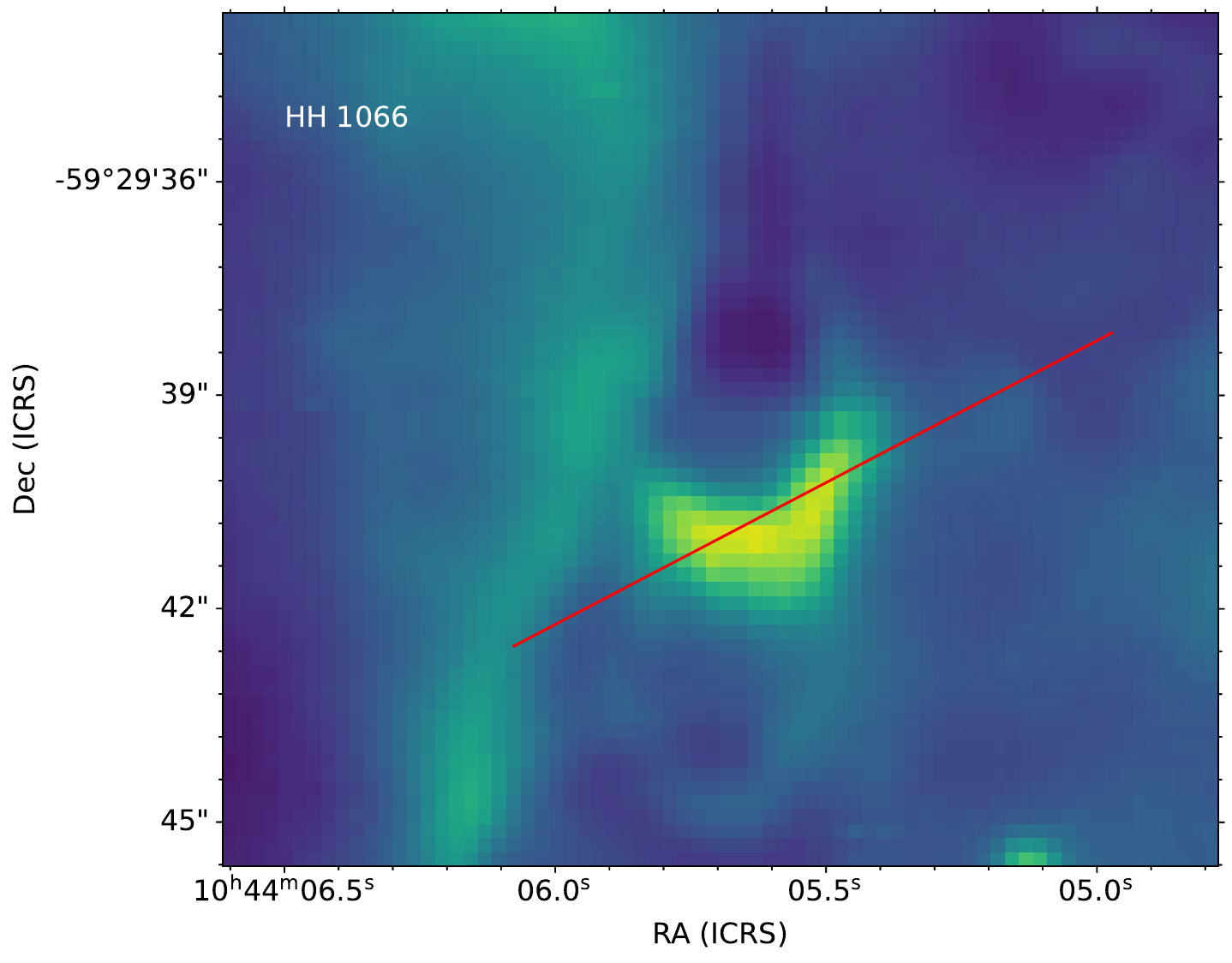}
    \includegraphics[width=\columnwidth]{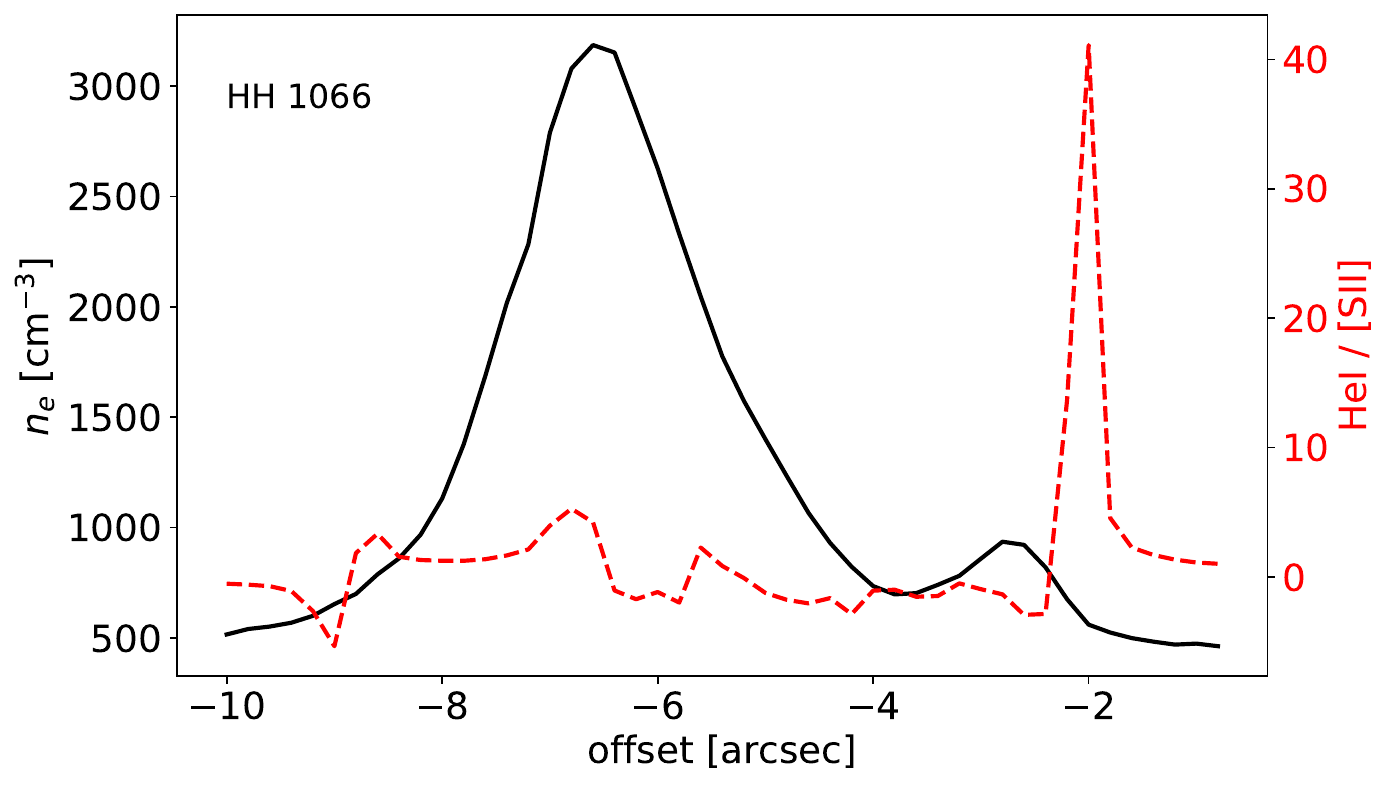}
    \includegraphics[width=\columnwidth]{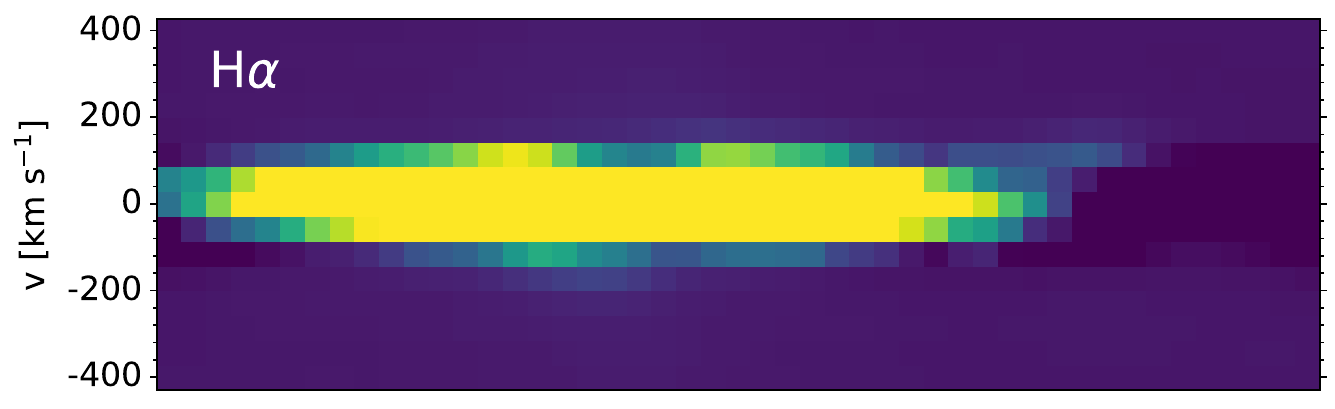}
    \includegraphics[width=\columnwidth]{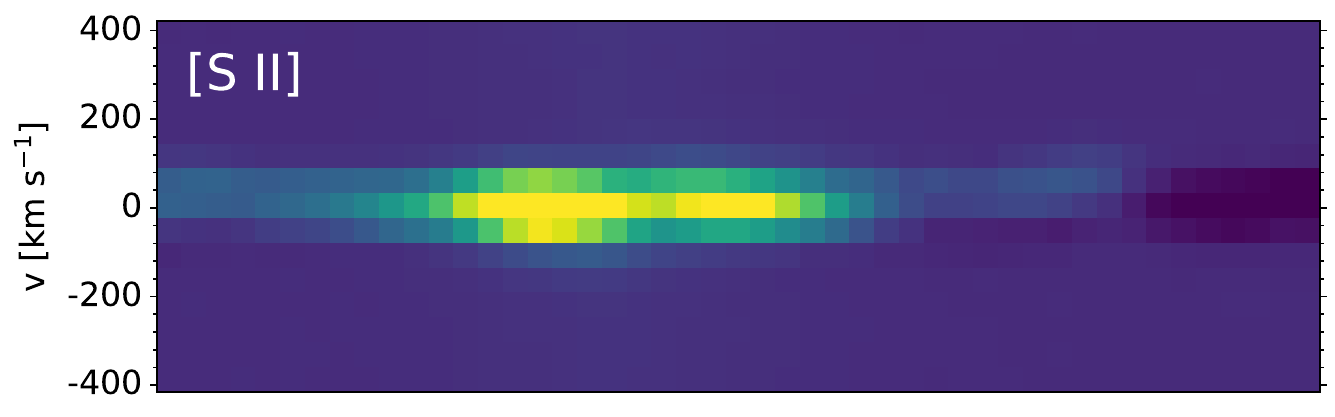}
    \includegraphics[width=\columnwidth]{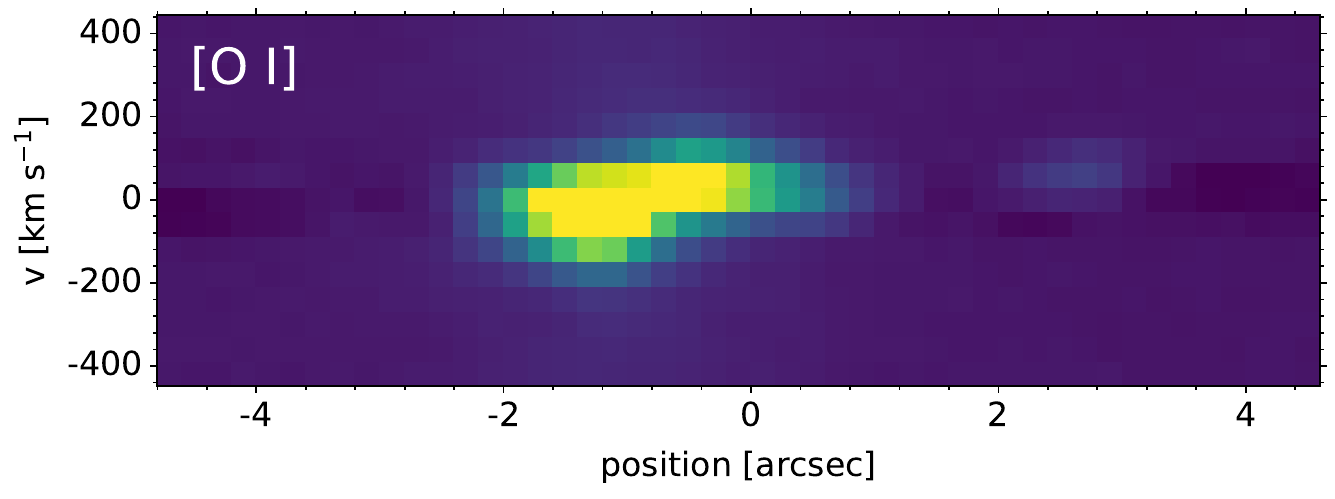}
    \includegraphics[width=\columnwidth]{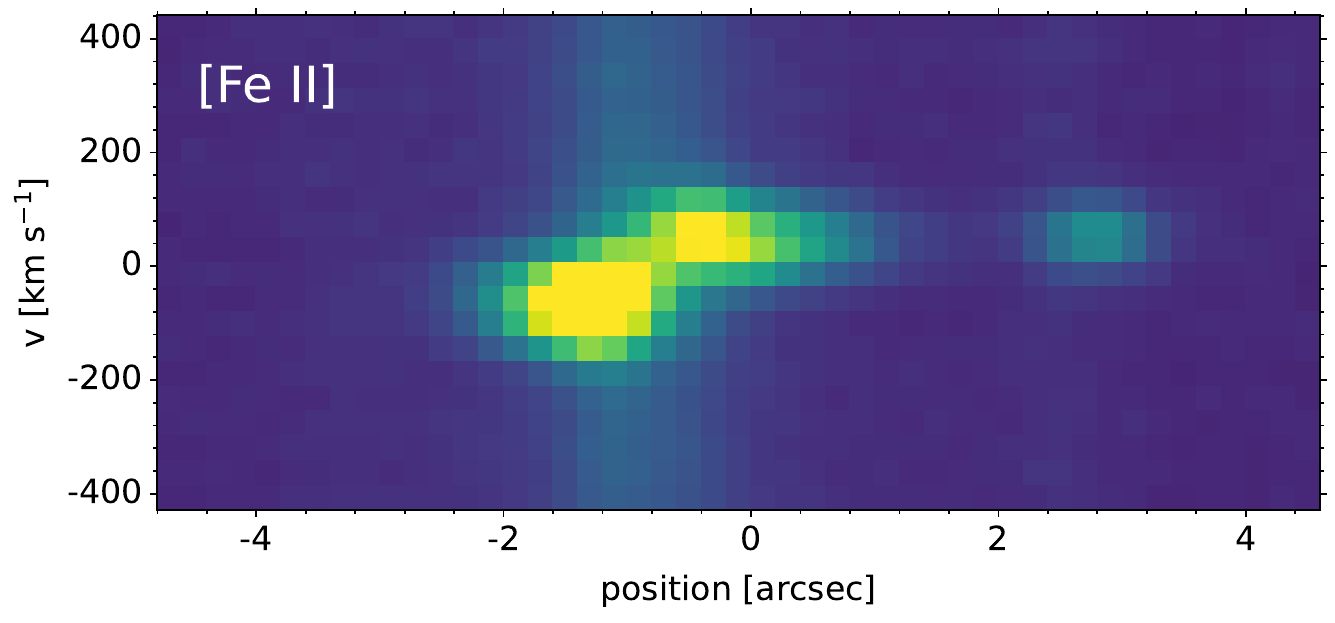}
    \includegraphics[width=\columnwidth]{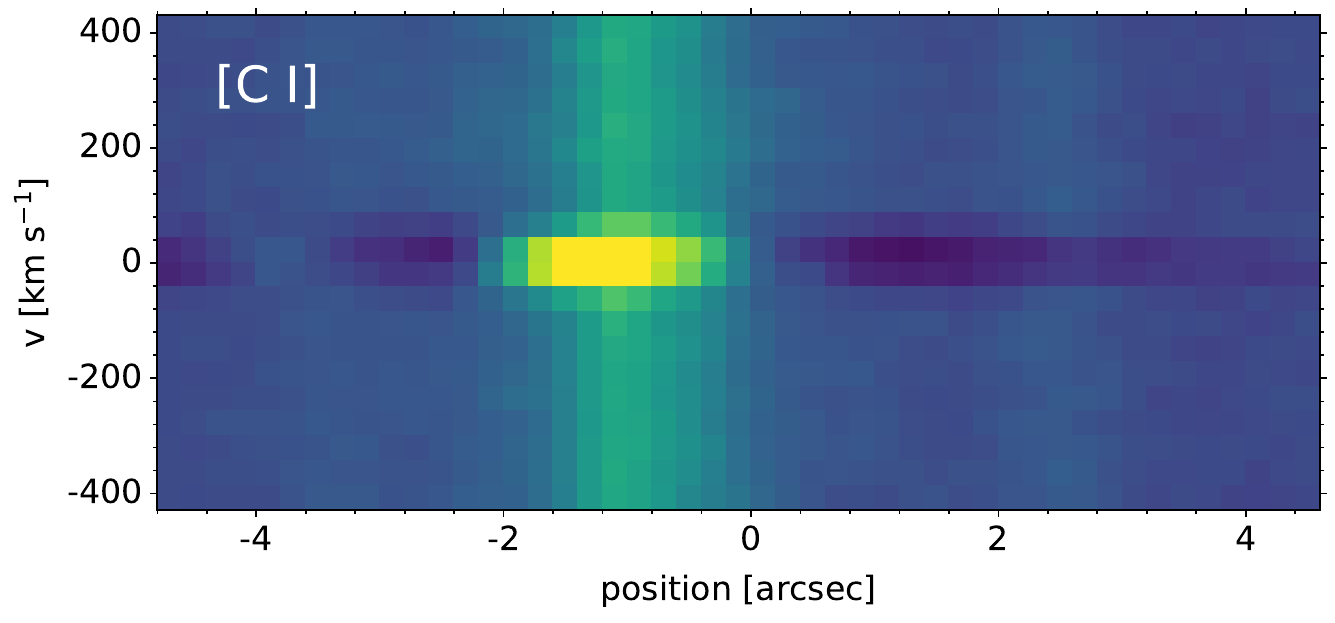}
    \caption{ Same as Figure~\ref{fig:hh901_jet_vels} but for HH~1066. Continuum emission from the driving source can be seen in the longer wavelength P-V diagrams ([Fe~{\sc ii}], [C~{\sc i}]). }
    \label{fig:hh1066_jet_vels}
\end{figure*}

\begin{figure}
	\includegraphics[width=\columnwidth]{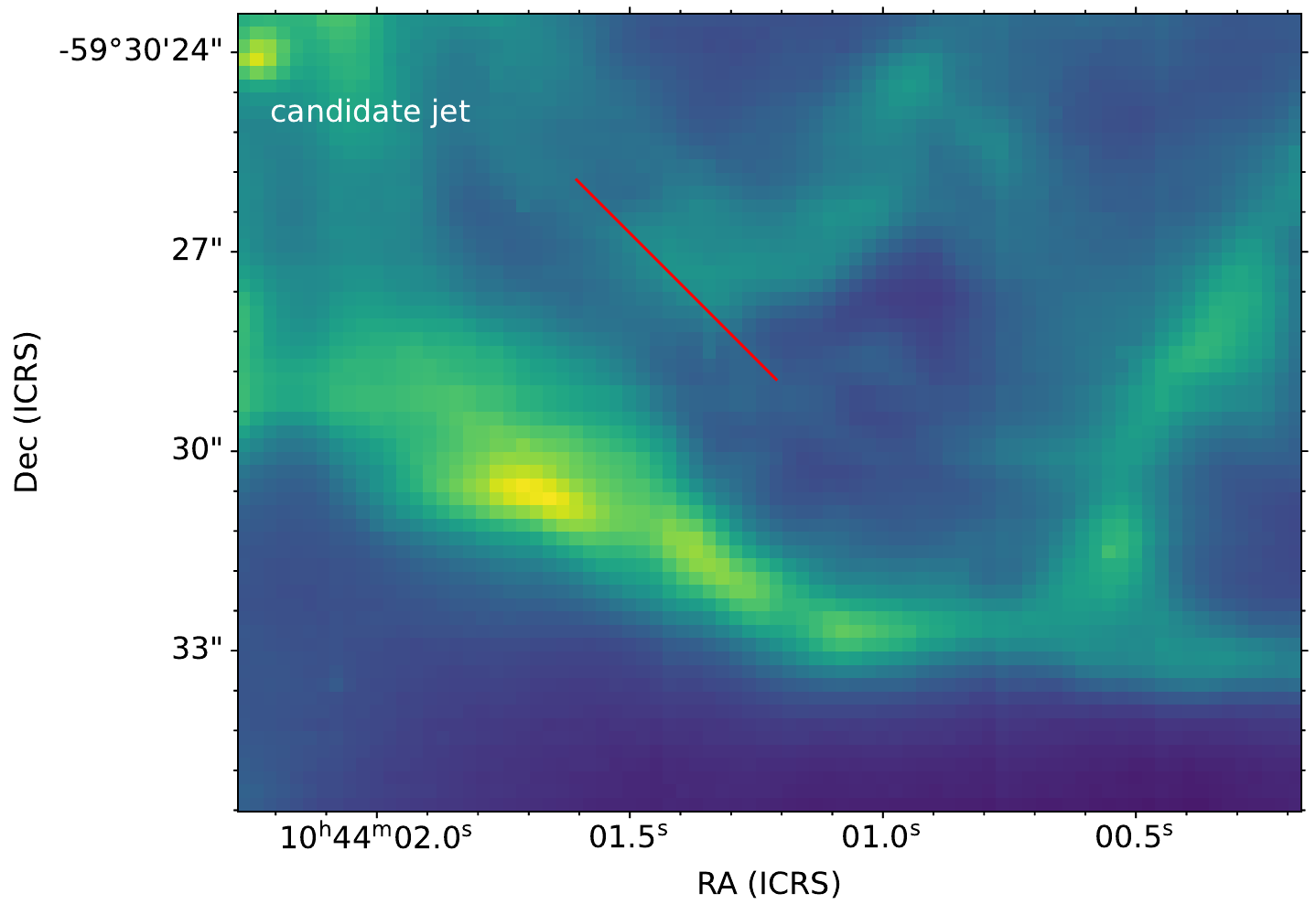}
    \includegraphics[width=\columnwidth]{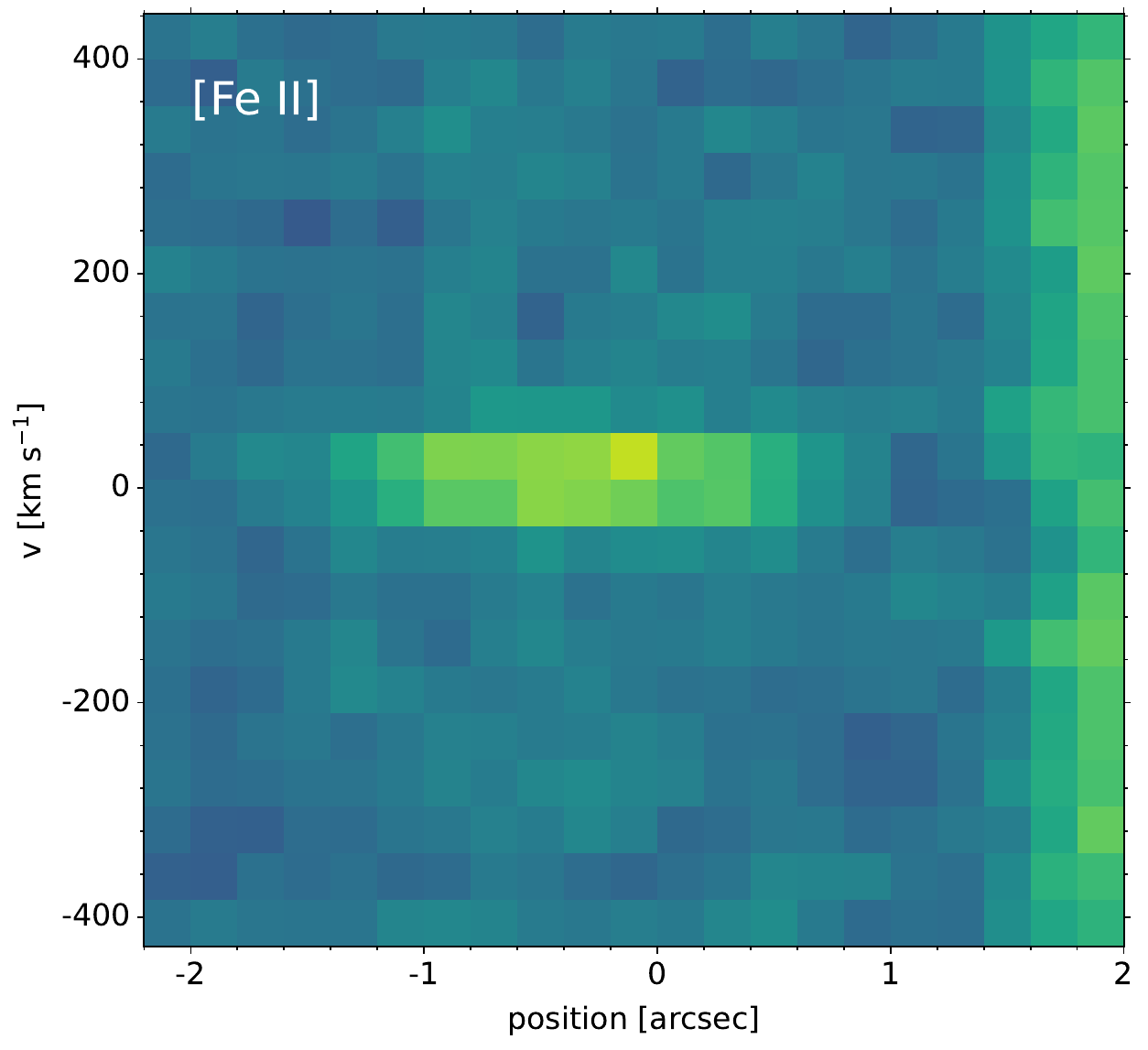}
    \caption{Same as Figure~\ref{fig:hh901_jet_vels} but for the candidate jet located above HH~902. We show only the [Fe~{\sc ii}] P-V diagram as this candidate jet can only be identified in the [Fe~{\sc ii}] lines. }
    \label{fig:cnj_jet_vels}
\end{figure}


\bsp	
\label{lastpage}
\end{document}